\documentclass[10pt,amsmath,amssymb,nofootinbib,twoside,twocolumn,superscriptaddress,floats,floatfix,aps,prd,preprintnumbers]{revtex4-2}
\usepackage[british]{babel}
\usepackage{xcolor}
\usepackage{booktabs}
\usepackage{latexsym}
\usepackage{amssymb}
\usepackage{amsmath}
\usepackage{graphicx}
\usepackage{tabularx}
\usepackage{cancel}
\usepackage{hyperref}
\usepackage{bm}
\usepackage{aas_macros}
\hypersetup{
    colorlinks=true,
    linkcolor=black,
    citecolor=blue,
    filecolor=black,
    urlcolor=blue,
    pdfauthor={Ivan~De~Martino},
    pdftitle={Dynamics of Dwarf Galaxies in Scalar-Tensor-Vector-Gravity}
}
\usepackage[capitalise]{cleveref}
\usepackage[utf8]{inputenc}
\usepackage[normalem]{ulem}
\usepackage{bm}
\begin{document}
\newcommand{\Salamanca}{\affiliation{Departamento de F\'isica Fundamental, Universidad de Salamanca, P. de la Merced, s/n, E-37008 Salamanca, Spain}}
\newcommand{\IUFFyM}{\affiliation{Instituto Universitario de Física Fundamental y Matemáticas (IUFFyM), P. de la Merced, s/n, E-37008 Salamanca, Spain}}

\author{Ivan~De~Martino}
\email{ivan.demartino@usal.es}
\Salamanca
\IUFFyM

\title{Dynamics of Dwarf Galaxies in Scalar-Tensor-Vector-Gravity}

\begin{abstract}
We have investigated whether the Scalar-Tensor-Vector Gravity theory (STVG) may explain the kinematic of stars in dwarf spheroidal galaxies. STVG modifies General Relativity by adding extra scalar and vector fields with the main aim of replacing dark matter in astrophysical self-gravitating systems. The weak-field limit of STVG brings a Yukawa-like modification to the Newtonian gravitational potential. The modification is modulated by two parameters, $\alpha$ and $\mu$, that represent a redefinition of the gravitational coupling constant and the mass of the additional vector fields, respectively.  Thus, adopting the modified gravitational potential arising in the weak-field limit of STVG, we have solved the spherical Jeans equation to predict the line-of-sight velocity dispersion profiles of eight dwarf spheroidal galaxies orbiting around the Milky Way. The predicted profiles are then compared to the data using a Monte Carlo Markov Chain algorithm. Our results pointed out some tensions on the $\alpha$ parameter within the data set, while comparison with previous analysis shows the effectiveness of STVG in replacing dark matter  with extra massive fields. Further improvements will require more sophisticated modelling of the line-of-sight velocity dispersion which will be possible as soon as high-precision astrometric data in dwarf spheroidals will become available.
\end{abstract}

\maketitle



\section{Introduction}

Nowadays, the $\Lambda$ Cold Dark Matter ($\Lambda$CDM) model is well established as the  {\em concordance} cosmological model. It provides an effective theoretical framework whose six parameters  are known with unprecedented accuracy \cite{Planck2020-V,Planck2020-VI,Planck2020-VII}. Although the framework is well consolidated some tensions with observations remain \cite{Abdalla2022,Peebles2022}, including, for instance, the well-known Hubble constant tension  \cite{DiValentino2021, Dainotti2023,Califano2022}. Another important puzzling issue of the {\em concordance} cosmological model is related to the fundamental nature of Cold Dark Matter (CDM) and its astrophysical predictions. Indeed, on the scale of galaxies, the CDM 
 predicts a steep mass density profile in the  core of each virialized dark matter halo which, however, disagrees with the observations of dwarf and low surface brightness galaxies that show a cored mass density profile.  This discrepancy is known as the cusp/core problem, and it is one of the well-known {small-scale} problems of the CDM paradigm  (for comprehensive reviews see, e.g., \cite{Boylan_Kolchin2011, Bullock2017,DelPopolo2017,deMartino2020,Salucci2021, Boldrini2021}) which are still unsolved. One way to resolve those small-scale challenges within the CDM framework relies on invoking baryonic feedback, but their efficiency is still debated  \cite{deMartino2020, Boldrini2021}. {Indeed, baryonic feedback is expected to be important down to stellar masses of the order of $\sim 10^{7.2}M_\odot$. However, a recent study used 
 tidal stability arguments of ultra-diffuse dwarf galaxies
in the Fornax Cluster to show that the cusp/core problem extends down
to $M_* \sim 10^{6}M_\odot$ \cite{Asencio2022}, and seems to favour Modified Newtonian Dynamics (MOND) over the $\Lambda$CDM model.}
 
 Hence, another possibility relies on modifying the underlying gravitational theory \cite{deMartino2020}.   
 Several modified theories of gravity have been  proposed and used to account for the dark matter content of the Universe. For instance, MOND successfully explains the kinematics of stars in galaxies \cite{Angus2008,Cardone2011,Angus2012,Angus2014,Angus2015,Chakrabarty2022} while it faces several challenges on cosmological scales (for a comprehensive review on MOND we refer to \cite{Famaey2012,Banik2022}). $f(R)$-gravity can explain the dynamics of galaxies and clusters of galaxies without resorting to dark matter \cite{Capozziello2009,Capozziello2012,Napolitano2012,Stabile2013,DeMartino2014,DeMartino2016,deMartino2023}, although it is unclear whether it can solve the small-scale problems of the CDM model \cite{deMartino2020}. Recently,  the Degenerate Higher-Order Scalar Tensor (DHOST) and the Refracted Gravity models  have been successfully tested in ultra-diffuse galaxies \cite{Laudato2022a, Laudato2022} and  in elliptical galaxies \cite{Cesare2020,Cesare2022}, respectively. Finally, tests of the screening mechanisms have been carried out showing the capability of screened modified theories of gravity of explaining the dynamics of galaxies in galaxy clusters \cite{Pizzuti2022a,Pizzuti2022b} {consistently with Solar System ephemerides.}.

Among these theories, the Scalar-Tensor-Vector Gravity theory (STVG), {sometimes referred to as MOdified Gravity (MOG)}, extends General Relativity (GR) by adding extra scalar and vector fields \cite{Moffat2006}. The main aim is to provide an alternative description of dark matter phenomenology. Indeed, massive vector fields are coupled to matter and exert a repulsive gravitational force on test particles which do not move along geodesics anymore. Then, the gravitational force is reduced with respect to the Newtonian one on small astrophysical scales, {\em i.e.} galactic and sub-galactic ones. In fact, the weak-filed limit of STVG theory shows a Yukawa correction to the Newtonian potential that is modulated by two parameters: $\alpha$ which represents the strength of the gravitational force, and $\mu$ which represents the inverse of the characteristic  Yukawa length  \cite{Moffat2013a, Moffat2014}. STVG has been widely tested in different astrophysical scenarios. Recently,  the orbital motion of the S2 star around the supermassive black hole at the centre of the Milky Way has been used to set an upper limit on the value of the strength parameter: $\alpha \leq 0.410$ at 99.7\%  confidence level \cite{DellaMonica2022, DellaMonica2023} which agrees with the prediction of the STVG for a gravitational mass source of $\sim 10^6 M_\odot$.
The STVG's parameters $\alpha$  and $\mu$ have also been bounded using galactic kinematics. For instance, the rotation curves extracted from the HI Nearby Galaxy Survey catalogue of galaxies were used to constrain $\alpha = 8.89 \pm 0.34$ and $\mu = 0.042 \pm 0.004$ kpc$^{-1}$ as averaged values of the whole sample \cite{Moffat2013a}. Furthermore, STVG has been satisfactorily tested with N-body simulations to study the global stability of a self-gravitating disc \cite{Ghafourian2017,Ghafourian2020}, and it has been used to  successfully fit multi-wavelengths observations of galaxy clusters,  and the direct detection of gravitational waves  \cite{Moffat2013, Moffat2014, Moffat2016, DeMartino2017}. On the other hand, analyses based on the dynamics of stars in dwarf spheroidal (dSph) and  low surface brightness ultra-diffuse galaxies have pointed out some inconsistencies. First, despite the fact that the masses and luminosities of the dSph galaxies  were comparable to each other, no shared values of $\alpha$ and $\mu$ were found within the data set \cite{Haghi2016}. Nevertheless, the lack of a proper statistical analysis and the use of simplified theoretical modelling may be the source of such an issue. Other inconsistencies arise in {\em (i)} low surface brightness ultra-diffuse galaxies \cite{deMartino2020b} where the dynamics of stars cannot be easily explained within STVG;  {\em (ii)} dwarf galaxies where STVG needs values of the observed stellar mass-to-light ratio higher than the one predicted by the stellar population synthesis models \cite{Bell2001} to correctly predict the rotation curves \cite{Haghighi2017}; {{\em (iii)} the velocity dispersion profile of the ultra-diffuse galaxy Dragonfly 44 ruled out STVG at  ruled out at 5.5$\sigma$ confidence \cite{Haghi2019}; and, finally, due to the lack of a fundamental acceleration scale, STVG cannot match the observed radial acceleration relation, making rotation curve of Milky Way is in strong tension with it \cite{Negrelli2018}.} 

Here, we use the internal dynamics of dSphs to constrain the parameters $\alpha$ and $\mu$.  We resolve the Jeans equation under the assumption of spherical symmetry (Sect.~\ref{sec:jeans_model})  to predict the velocity dispersion profiles and fit the theoretical prediction to the data employing a Monte Carlo Markov Chain (MCMC) algorithm (Sects.~\ref{sec:data} and \ref{sec:results}). Afterwards, we provide an interpretation of our results comparing them  with those obtained  in spiral galaxies \cite{Moffat2013a}, in dSphs \cite{Haghi2016}, in ultra-diffuse galaxies \cite{deMartino2020b} and in galaxy clusters \cite{DeMartino2017}. Finally, in Sect.~\ref{sec:discussions}, we give our conclusion and future perspectives. 

\section{A brief introduction to the weak-field limit of STVG} \label{sec:Yukawa}

Here we will give a brief summary of the main steps leading to the weak-field approximation of the STVG  \cite{Moffat2006,Moffat2013a,Moffat2014,Banerjee2017}. Let us start with the action  which can be decomposed as the sum of four terms: $S=S_{EH}+S_{\phi}+S_s+S_m$;  here, $S_{EH}$ is the Einstein-Hilbert action of GR with a cosmological constant, $S_{\phi}$ expresses the contribution of the massive vector field $\phi^{\rho}$, $S_s$ is the contribution of the scalar fields $G$ and $\mu$ and, finally, $S_m$ encodes the matter field. In more detail one has 
\begin{align}
S_{G}& =\frac{1}{16\pi}\int\frac{1}{G}\left(R-2\Lambda\right)\sqrt{-g}d^4x
\label{11a}\,,\\
S_{\phi}&=-\frac{1}{4\pi}\int\omega \mathcal{K} \sqrt{-g}d^4x\,,\\
S_{s}&=-\int\frac{1}{G}\left[\frac{1}{2}g^{\rho\tau}\mathcal{G}_{\rho\tau}
+\frac{V_G(G)}{G^2}+\frac{V_\mu(\mu)}{\mu^2}\right]\sqrt{-g}d^4x\,,
\end{align}
where
\begin{align}
    \mathcal{K} &= \frac{1}{4}B^{\rho\tau}B_{\rho\tau}-\frac{1}{2}\mu^2\phi^\rho\phi_\rho+V_{\phi}\left(\phi_{\rho} \phi^{\rho}\right)\,,\\
    B_{\rho\tau}&=\partial_\rho\phi_\tau-\partial_\tau\phi_\rho\,,\\
    \mathcal{G}_{\rho\tau}&= \frac{\nabla_\rho G \nabla_\tau G}{G^2}+\frac{\nabla_\rho \mu \nabla_\tau \mu}{\mu^2}\,.
\end{align}

Here, $\phi^{\rho}$ is the vector field whose mass $\mu$ is a scalar field  and the gravitational constant $G$ is upgraded to a scalar field. Finally, $\omega$ is the dimensionless coupling constant and  $V_\phi$, $V_G$, and $V_\mu$ are the self-interaction potentials for the vector and scalar fields, respectively. Since the matter is  coupled to the massive vector field,  a fifth force arises and does not allow particles to follow geodesics. Indeed, the equation of motion for a test particle in STVG is \cite{Moffat2006,Moffat2013a}
\begin{equation}
m\left(\frac{d^2x^\nu}{ds^2}+\Gamma^{\nu}_{\rho\tau}\frac{dx^\rho}{ds}\frac{dx^\tau}{ds}\right)=\lambda\omega B^{\nu}_{\ \zeta} \frac{dx^\zeta}{ds}
\label{1e}
\end{equation} 
where $s$ is the affine parameter along the trajectory, and $\lambda$ is a coupling constant related to the mass of the particle: $\lambda=\kappa m$. 
By taking the spatial divergence of Eq. \eqref{1e}  one obtains  \cite{Moffat2013a}
\begin{equation}
\bm{\nabla}\cdot\bm{g}-\frac{1}{2}\nabla^2h_{00}=-\omega\kappa\nabla^2\phi_0\,.
\label{geo}
\end{equation}

To solve the above equation, we need an expression for the $\nabla^2\phi_0$ term.
By varying the action with respect to the vector field one obtains  the following field equation for the massive vector field \cite{Moffat2006}
\begin{equation}
\nabla_\rho B^{\rho\tau}-\mu^2\phi^\tau=-\frac{4\pi J^\tau}{\omega}\,,
\label{1b}
\end{equation}
where 
\begin{equation} 
J^\tau=-\frac{1}{\sqrt{-g}}\frac{\delta S_m}{\delta\phi^\tau}\,,
\end{equation}
and it encodes the coupling to the matter field. Linearising Eq. \eqref{1b} with respect to a Minkowski background space-time, one  gets \cite{Moffat2013a}
\begin{equation}\label{vec}
\nabla^2\phi_0-\mu^2\phi_0=-\frac{4\pi J^0}{\omega}\,.
\end{equation}

Under the assumption that the density of the massive vector field is smaller than the density of the matter field, $\mu$ and $G$ turn out to be constant. Hence,  Eq. \eqref{vec} and \eqref{geo} provides the modified Poisson equation that can be solved to obtain the modified gravitational potential \cite{Moffat2013a}:
\begin{align}
\Phi\left(\textbf{r}\right)=&-G_N\left(1+\alpha\right)\int\frac{\rho\left(\textbf{r}'\right)}{\mid\textbf{r}-\textbf{r}'\mid}d^3r'\nonumber\\
 &+G_N\alpha\int\frac{\rho\left(\textbf{r}'\right)}{\mid\textbf{r}-\textbf{r}'\mid}e^{-\mu\mid\textbf{r}-\textbf{r}'\mid}d^3r'\,.
\label{1f}
\end{align}
Here $\alpha=\frac{G_{\infty}-G_N}{G_N}\geq 0$, $G_N$ and $G_\infty$ are the Newtonian gravitational constant and effective gravitational constant at infinity, respectively. 
Let us note that the first term in Eq. \eqref{1f} is the usual attractive term of Newtonian gravity whose gravitational constant (or, alternatively,  gravitational mass) is enhanced by a factor $(1+\alpha)$, while the second term provides a repulsive fifth force enhanced by a factor $\alpha$ but also modulated by the Yukawa term $e^{-\mu\mid\textbf{r}-\textbf{r}'\mid}$. The astrophysical effects of dark matter could be then ascribable to such a repulsive term.

Particularizing the modified gravitational potential in Eq. \eqref{1f} to the case of a spherically symmetric matter distribution, one can easily obtain the radial acceleration 
\begin{eqnarray}
a\left(r\right)&=&-\frac{d\Phi}{dr}\nonumber\\
&=& -\frac{4\pi G_N\alpha}{\mu r^2}\biggl\{\frac{1+\alpha}{\alpha}I_1(r)-\left(1+\mu r\right)e^{-\mu r} I_2(r)\nonumber \\&&-\left[\sinh\left(\mu r\right)-\mu r \cosh\left(\mu r\right)\right]I_3(r)\biggr\}
\label{STVG}
\end{eqnarray}
where
\begin{align}
    I_1 (r) &=\int_{0}^{r}r'^2\rho(r')dr'\,,\\
    I_2 (r) &=\int_{0}^{r}r'\rho(r')\sinh\left(\mu r'\right)dr'\,,\\
    I_3 (r) &=\int_{r}^{\mathcal{R}}r'\rho(r')e^{-\mu r'}dr'\,.
\end{align}

Here, $\mathcal{R}$ determines the physical size of the system. We remark that, since STVG aims to replace dark matter with the scalar and vector fields, the mass density $\rho(r)$ appearing in the previous equations coincides with the stellar mass density profile  $\rho_*(r)$ that we will introduce in the following section.

\section{Jeans analysis}\label{sec:jeans_model}

The gravitational potential well fully determines the stellar kinematics of a self-gravitating system  in dynamical equilibrium and supported by the velocity dispersion. 
Let us reduce ourselves to the case of a spherically symmetric dwarf galaxy to shape the kinematics of the stars through the spherical Jeans equation \cite{Lokas2003,Mamon2005,Binney2008,Mamon2010}  
 \begin{equation}\label{eq:Jeans}
 	\frac{d[\rho_*(r)\sigma_r^2(r)]}{dr} + 2{  \beta}\frac{\rho_*(r)\sigma_r^2(r)}{r} = -\rho_*(r)\frac{d\Phi(r)}{dr}\,.
 \end{equation}

Here $\frac{d\Phi(r)}{dr}$ is the modified gravitational acceleration in Eq. \eqref{STVG}, $\sigma_r(r)$ is the radial component of the velocity dispersion, $\rho_*(r)$ is the mass density profile of the tracing stellar population and, finally, ${ \beta}$  is the velocity anisotropy parameter that, hereby, will be considered to be a constant. In such a case, Eq. \eqref{eq:Jeans} has the following solution  \cite{Lokas2003} 
 \begin{equation}
 	\rho_*(r)\sigma_r(r)=r^{-2{  \beta}}\int_{r}^{\infty} \frac{d\Phi(x)}{dx}\rho_*(x)x^{2{  \beta}} \,dx \,.
 \end{equation}
Nevertheless, it is worth noting that $\sigma_r(r)$ must be projected along the line of sight in order to be fitted to the data. The projected velocity dispersion, $\sigma^2_{\mathrm{los}}$ is then   \cite{Binney2008}
\begin{equation}\label{eq:sigmalos}
	\sigma^2_{\mathrm{los}} (R_{\rm p}) = \frac{2}{\Sigma_*(R_{\rm p})}\int_{R_{\rm p}}^{\infty}\biggl(1-{  \beta}\frac{R_{\rm p}^2}{r^2}\biggr) \frac{\sigma^2_r(r)\rho_*(r)}{(r^2-R_{\rm p}^2)^{1/2}} r \,dr\,,
\end{equation}
where $R_{\rm p}$ is the galactic radius projected onto the sky and, finally, $\Sigma_*(R_{\rm p})$ is the stellar surface mass density. {The latter is set to the Plummer profile and can be derived by the three-dimensional mass density profile 
\begin{equation}
	\rho_*(r) =  { \frac{M_*}{L_V}}\frac{3 L_{V}}{4\pi r_{1/2}^3}\left(1+\frac{r^2}{r_{1/2}^2}\right)^{-\frac{5}{2}}\,,
	\label{Plummer}
\end{equation}
once the previous equation is projected, hence resulting in
\begin{equation}\label{eq:surfacePlummer}
    \Sigma_*(R_{\rm p}) = { \frac{M_*}{L_V}} \frac{L_V}{\pi r_{1/2}^2}\left(1+\frac{R_{\rm p}^2}{r_{1/2}^2}\right)^{-2}\,.
\end{equation}
}

The Plummer profile is fully determined once the luminosity in the $V$-band ($L_V$), the stellar mass-to-light ratio (${M_*}/{L_V}$), and the half-light radius ($r_{1/2}$) are measured or estimated from observations \cite{Walker2009d}.

\begin{table*}
		\begin{center}
			\resizebox{14cm}{!}{
				\setlength{\tabcolsep}{4pt}
				\begin{tabular}{|lcccccc|}
					\hline
					Galaxy & $D_{\odot}$ &  $D_{p}$ &$\log(L_{\rm V})$  & $r_{1/2}$ & $M_*/L_V$  &Ref. \\
					& (kpc)                & (kpc)                &($L_{\odot}$)      & (pc)     &    &  \\
					(1)         & (2)      & (3)  & (4) & (5)  & (6) &  (7) \\
					\hline
					\textbf{Carina} & 105$\pm$6  & $60^{+21}_{-16}$ & 5.57$\pm$0.20 & 273$\pm$45 & $3.4\pm2.9$  & \cite{Pietrzynski2009,Irwin1995,Walker2009c,Fritz2018,deMartino2023}  \\[0.1cm]
					\textbf{Draco} & 76$\pm$5  & $28^{+12}_{-7}$  &5.45$\pm$0.08 & 244$\pm$9 & $11.1\pm4.7$ &  \cite{Walker2009c,Bonanos2004,Martin2008,Walker2007,Fritz2018,deMartino2023}\\[0.1cm]
					\textbf{Fornax} & 147$\pm$12  & $69^{+26}_{-18}$  &7.31$\pm$0.12 & 792$\pm$58&  $7.1\pm6.0$ & \cite{Pietrzynski2009,Irwin1995,Walker2009c,Fritz2018,deMartino2023} \\[0.1cm]
					\textbf{Leo I} & $254^{+19}_{-16}$  & $45^{+80}_{-34}$ & 6.74$\pm$0.12 & 298$\pm$29& $8.8\pm5.6$&  \cite{Irwin1995,Walker2009c,Bellazzini2004,Mateo2008,Fritz2018,deMartino2023}  \\[0.1cm]
					\textbf{Leo II} & 233$\pm$15  & $45^{+121}_{-30}$   & 5.87$\pm$0.12 & 219$\pm$52 & $0.4\pm0.4$  & \cite{Irwin1995,Walker2009c,Bellazzini2005,Koch2007,Fritz2018,deMartino2023}\\[0.1cm]
					\textbf{Sculptor} & 86$\pm$6  & $50^{+15}_{-10}$  &6.36$\pm$0.20 & 311$\pm$46 & $3.6\pm2.0$ & \cite{Irwin1995,Walker2009c,Pietrzynski2008,Fritz2018,deMartino2023} \\[0.1cm]
					\textbf{Sextans} & 86$\pm$4  & $71^{+11}_{-12}$  &5.64$\pm$0.20 & 748$\pm$66 & $8.5\pm3.3$ & \cite{Irwin1995,Walker2009c, Lee2009,Fritz2018,deMartino2023}\\[0.1cm]
					\textbf{Ursa Minor} & 76$\pm$4  & $29^{+8}_{-6}$  & 5.45$\pm$0.20 & 398$\pm$44 & $1.2\pm1.3$ &\cite{Irwin1995,Walker2009c,Carrera2002,Walker2009b,Fritz2018,deMartino2023}\\[0.1cm]
					\hline
				\end{tabular}
			}
		\end{center}
	\caption{Observational properties of the eight dSphs {analysed in this work}. Columns (2) and (3): distance of the dSph  from the observer and distance of the pericentre of the dSph orbit around the Milky Way from the Milky Way { center of mass}; Column (4): total $V$-band luminosity; Column (5): half-light radius; Column (6): the stellar mass-to-light ratio estimated by \cite{deMartino2023} using stellar population synthesis models in \cite{Bell2001}; and Column (6): references from which data were extracted.
}\label{tab:1}
	\end{table*}

\section{Data and data analysis}\label{sec:data}

In our analysis, we will predict the theoretical velocity dispersion profile projected along the line of sight by solving Eq. \eqref{eq:sigmalos} and fit it to the measured line-of-sight velocity dispersion profiles of the eight dSphs, namely Carina, Fornax, Sculptor,  Sextans, Draco, Leo I, Leo II, and Ursa Minor, reported in Table \ref{tab:1}. As a product of this procedure, for each galaxy, we will estimate the best-fit values of the STVG's parameters $\alpha$ and $\mu$, the velocity anisotropy parameter $\beta$, and the stellar mass-to-light ratio $M_*/L_V$, and their corresponding uncertainties using an MCMC analysis. 

In more detail, the kinematic data sets of the following dSphs: Carina, Fornax, Sculptor, and Sextans; were  obtained with the Michigan/MIKE Fiber Spectrograph \cite{Walker2007, Walker2009a, Walker2009b, Walker2009c, Walker2009d}. On the other hand, the kinematic data sets of Draco, Leo I, Leo II, and Ursa Minor were obtained 
with the Hectochelle fiber spectrograph at the MMT \cite{Mateo2008}. Additionally, for each galaxy, the values of the  luminosity in the $V$-band,  the stellar mass-to-light ratio, and the half-light radius  are taken from \cite{Pietrzynski2009,Irwin1995,Walker2009c,Bonanos2004,Martin2008,Walker2007,Bellazzini2004,Mateo2008,Bellazzini2005,Koch2007,Pietrzynski2008,Lee2009,Carrera2002,Walker2009b,Fritz2018} and listed in  Table \ref{tab:1}. Finally, following \cite{deMartino2023}, the physical size of the system $\mathcal{R}$ is set for each galaxy  as the radius where the mass density profile is decreased by 99\% w.r.t. the central density.

Generally speaking, the total mass-to-light ratio of a dSph depends on the mass of the dark matter halo. However in STVG dark matter is absent, therefore the mass-to-light ratio needed to fit the kinematic data sets is expected to be  the stellar mass-to-light ratio whose estimation is based on stellar population synthesis models \cite{Bell2001}. Following \cite{deMartino2023},  we still adopt $M_*/L$ as a free parameter but we will vary it according to the averaged values of $M_*/L$ shown in Table \ref{tab:1}.

\subsection{Methodology}\label{sec:mcmc}

Our modelling procedure predicts  the projected velocity dispersion profile $\sigma_{\mathrm{los,\, th}}(r)$  in STVG, and uses the projected velocity dispersion profile data sets $\sigma_{\mathrm{los,\, obs}}(r)$ measured by \cite{Walker2009d} with their observational uncertainties ($\Delta\sigma_{\mathrm{los,\, obs}}(r_i)$) to provide an estimation  of the best-fit values and their corresponding uncertainties for a set of four free parameters: $\bm{\theta} =$ \{$\alpha$, $\mu$, $\beta$, $M_*/L_V$\}. The four-dimensional parameter space is explored by employing the MCMC algorithm \texttt{emcee} \cite{emcee}.  We set a uniform prior distribution on $\log\alpha\in [-3;3]$, $\mu  \in (0, 10]\times(10^{-2} {\rm kpc}^{-1})$, and $\beta\in [-100, 1)$. Finally, for each dSph, we set a Gaussian prior on the stellar mass-to-light ratio $M_*/L_V$ with mean value and dispersion set according to Table \ref{tab:1} (those values are taken from Column (13) of Table 1 in \cite{deMartino2023}). Finally, the posterior probability  distribution is given by the following likelihood function
    \begin{align}
    	-2\log \mathcal{L}(\bm{\theta}|\textrm{ data}) \propto& \sum_i\biggl[\frac{\sigma_{\mathrm{los,\, th}}(\bm{\theta},\, R_{p,i})-\sigma_{\mathrm{los,\, obs}}(R_{p,i})}{\Delta\sigma_{\mathrm{los,\, obs}}(R_{p,i})}\biggr]^2\,,
    	\label{eq:likelihood}
    \end{align}
and we run 12 chains that we consider they have converged when the length of each chain is 100 times longer than the autocorrelation  time and the latter changes by less than 1\% (for more details we refer to Sec. 3 of \cite{deMartino2022}).

\section{Results} \label{sec:results}
\begin{table}
		\begin{center}
			\resizebox{\columnwidth}{!}{
				\setlength{\tabcolsep}{4pt}
				\begin{tabular}{|lcccc|}
					\hline
					Galaxy & $\log\alpha$ &  $\mu$ &$\beta$  &  $M_*/L_V$  \\
                                &          & ($10^{-2}$ kpc$^{-1}$)  & & \\
					(1)         & (2)      & (3)  & (4) & (5)   \\
					\hline
        &          &   & & \\[-0.3cm]
					\textbf{Carina} & $1.1^{+0.4}_{-0.3}$ & $0.08^{+1.60}_{-0.07}$ & $-1.7^{+0.7}_{-1.2}$ & $2.8^{+2.2}_{-1.8}$ \\[0.1cm]
					\textbf{Draco} & $0.9\pm 0.2$ & $0.07^{+1.60}_{-0.06}$ & $-14.1^{+5.6}_{-4.1}$ & $8.1^{+3.0}_{-2.9}$  \\[0.1cm]
					\textbf{Fornax} & $1.9^{+0.4}_{-0.2}$ & $0.10^{+1.93}_{-0.09}$ & $-0.26^{+0.09}_{-0.10}$ & $5.5^{+3.6}_{-3.1}$  \\[0.1cm]
					\textbf{Leo I} &  $1.0^{+0.3}_{-0.2}$ & $0.10^{+2.22}_{-0.09}$ & $-2.8^{+1.4}_{-4.2}$ & $6.5^{+3.5}_{-3.3}$ \\[0.1cm]
					\textbf{Leo II} & $1.5^{+0.4}_{-0.3}$ & $0.10^{+2.00}_{-0.09}$ & $-0.6^{+0.7}_{-2.1}$ & $0.5\pm 0.3$  \\[0.1cm]
					\textbf{Sculptor} & $1.7^{+0.3}_{-0.2}$ & $0.09^{+1.52}_{-0.08}$ & $-1.4^{+0.3}_{-0.4}$ & $2.8^{+1.4}_{-1.4}$   \\[0.1cm]
					\textbf{Sextans} &  $1.0\pm 0.2$ & $0.09^{+1.91}_{-0.08}$ & $-0.2\pm 0.2$ & $6.0\pm{2.2}$ \\[0.1cm]
					\textbf{Ursa Minor} & $2.0^{+0.4}_{-0.2}$ & $0.10^{+2.20}_{-0.09}$ & $-1.1^{+0.5}_{-0.7}$ & $1.2^{+0.9}_{-0.7}$ \\[0.1cm]
					\hline
				\end{tabular}
			}
		\end{center}
	\caption{The Table reports the median and the 68\% confidence intervals of the posterior distribution of the parameters  $\bm{\theta} = \{\alpha, \mu, \beta, M_*/L\}$ for all the dSphs.}\label{tab:2}
	\end{table}

We employed a MCMC algorithm to explore the four-dimensional parameter space  $\bm{\theta} = \{\alpha, \mu, \beta, M_*/L\}$, and to estimate the values of the parameters  $\bm{\theta}$ that can fit the observational data sets of the line-of-sight velocity dispersion profiles of the dSphs galaxies Carina, Draco, Fornax, Leo I, Leo II, Sculptor, Sextans, and Ursa Minor.  The median and the 68\% confidence intervals of the posterior distribution of the parameters  are reported in Table \ref{tab:2}. 

\begin{figure*}[ht]
\begin{center}
\begin{tabular}{c}
\includegraphics[width=0.99\columnwidth]{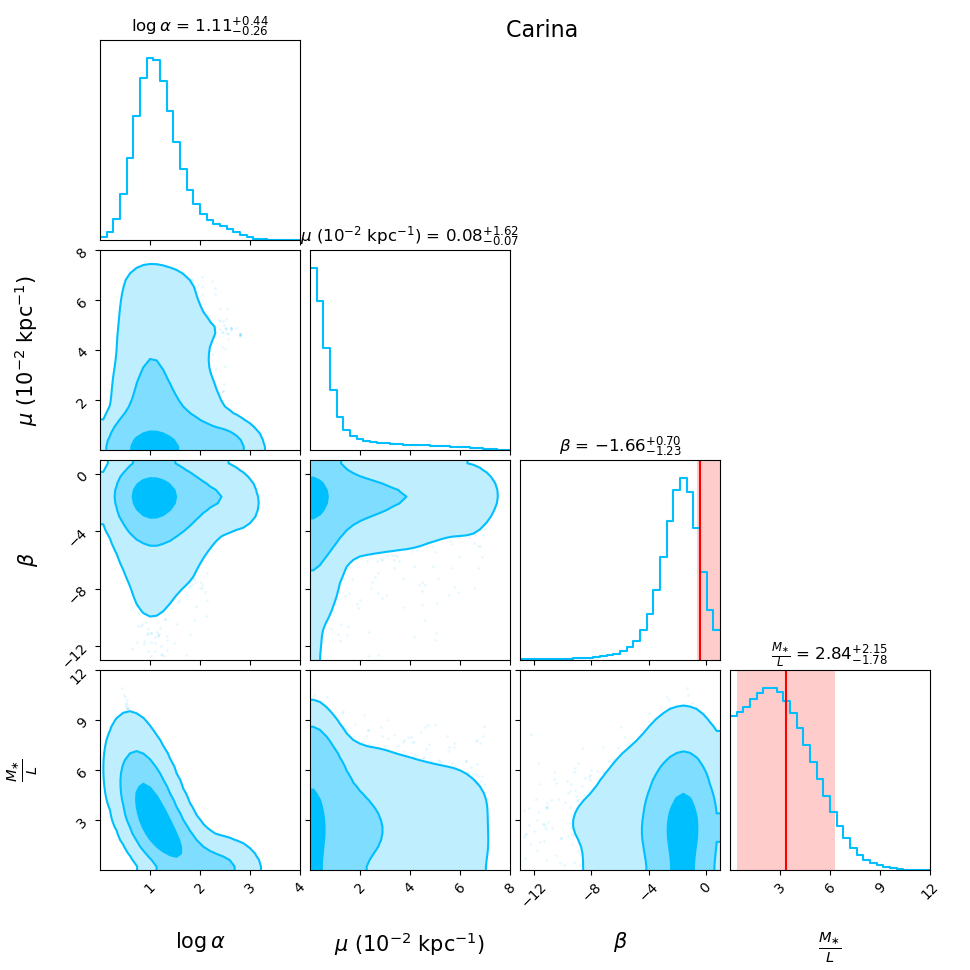}
\includegraphics[width=0.99\columnwidth]{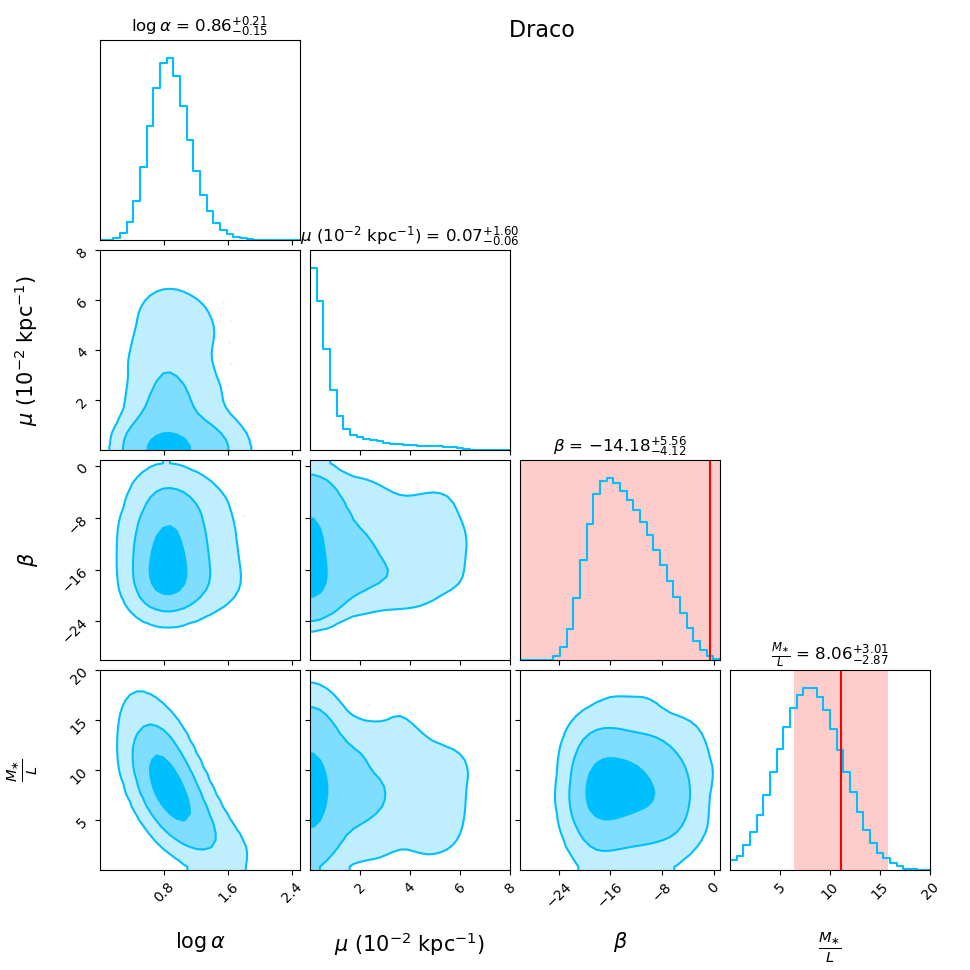}\\
\includegraphics[width=0.99\columnwidth]{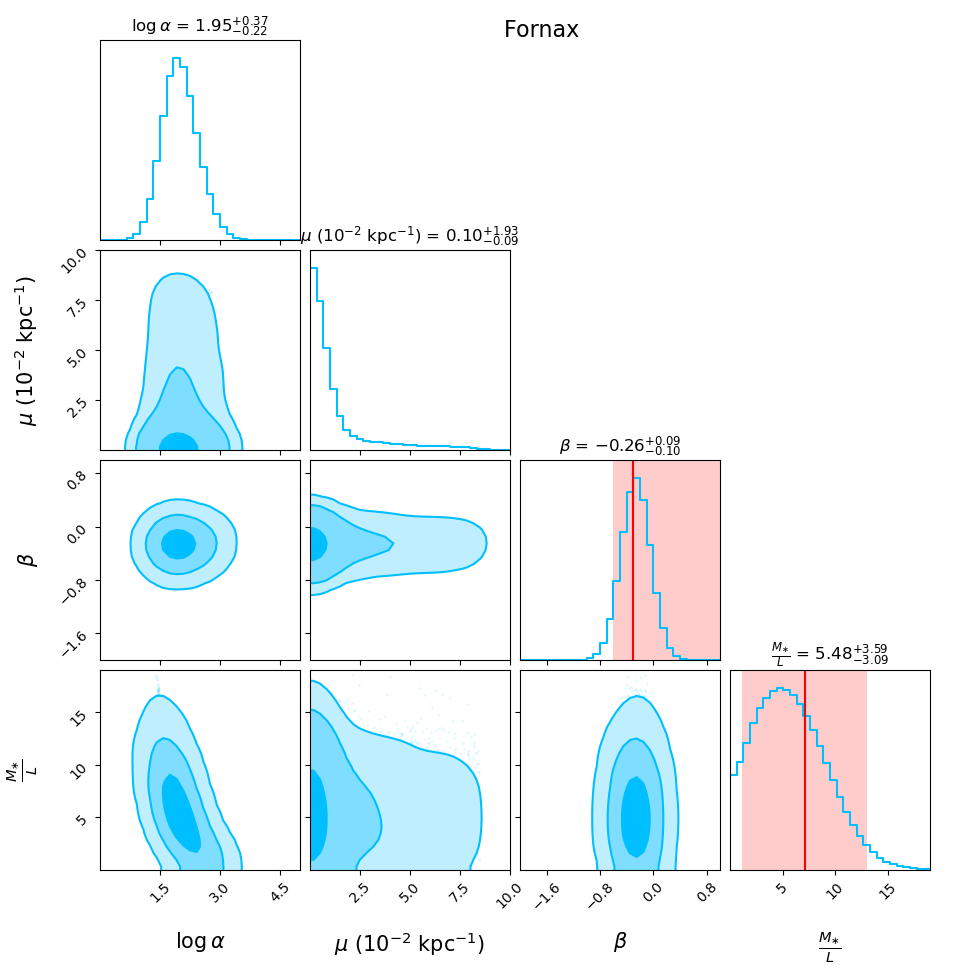}
\includegraphics[width=0.99\columnwidth]{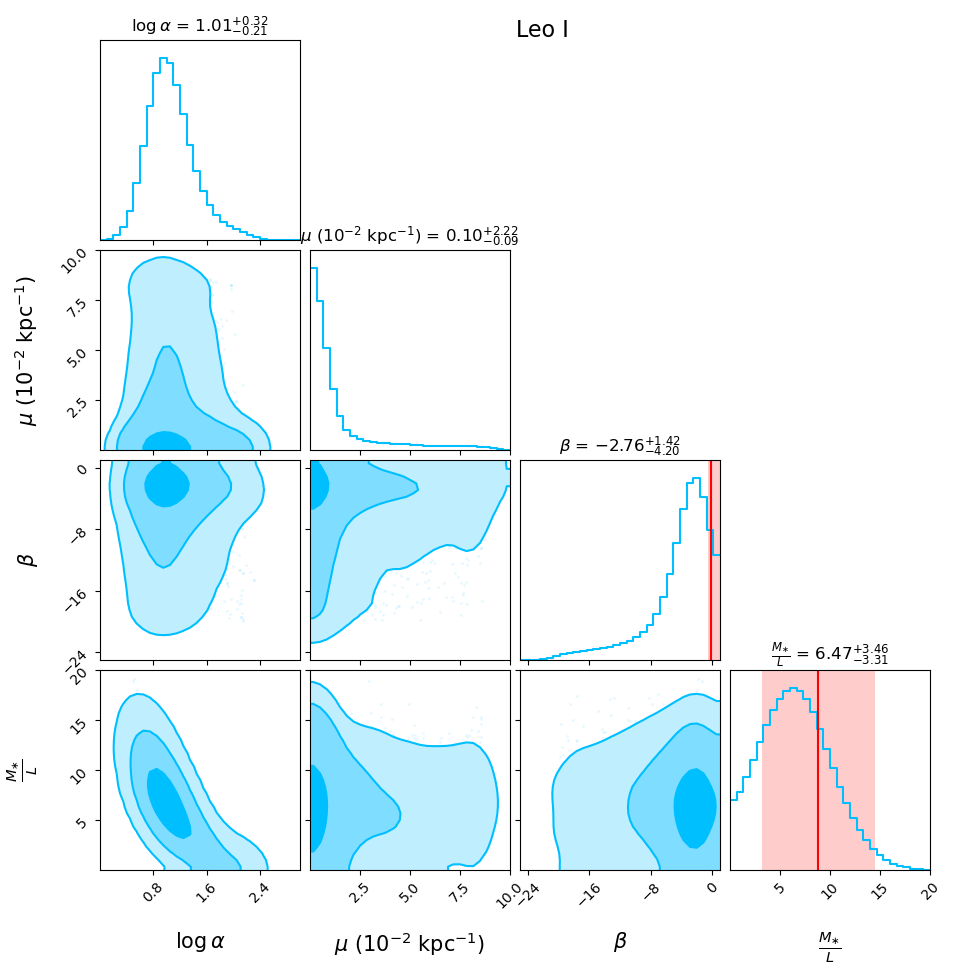}
\end{tabular}
\end{center}
\caption{MCMC posterior distributions of the parameters $\bm{\theta} = \{\log\alpha, \mu, \beta, M_*/L\}$ for Carina Draco, Fornax, and Leo I. The blue-shaded areas with decreasing darkness depict the 68\%, 95\%, and 99\% confidence regions, respectively. On top of each column, we report the median values of the posterior distributions with their 68\% confidence intervals.  The red shaded areas correspond to the best fit {values} and the $1\sigma$  {uncertainties}  of the velocity anisotropy parameter reported in \citet{Walker2009d}, and the expected values of $M_*/L$  listed in Table \ref{tab:1}.}
\label{fig:1}
\end{figure*}

\begin{figure*}[ht]
\begin{center}
\begin{tabular}{c}
\includegraphics[width=0.99\columnwidth]{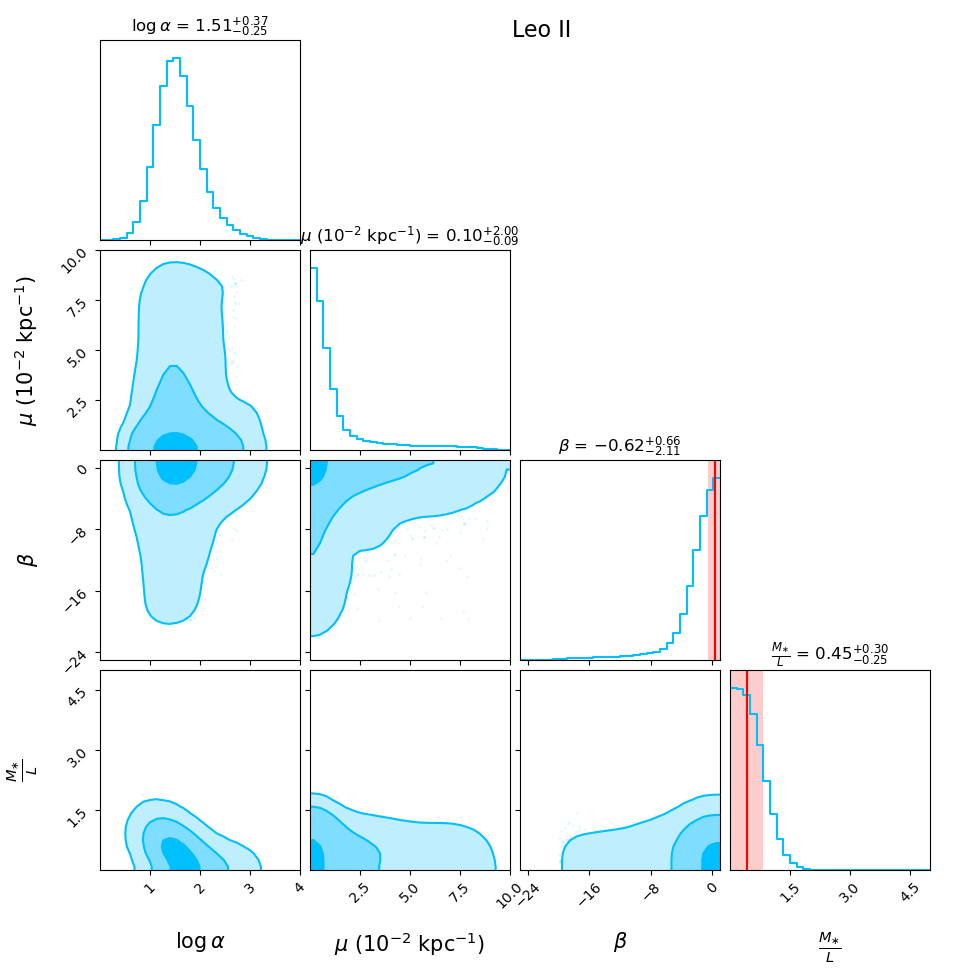}
\includegraphics[width=0.99\columnwidth]{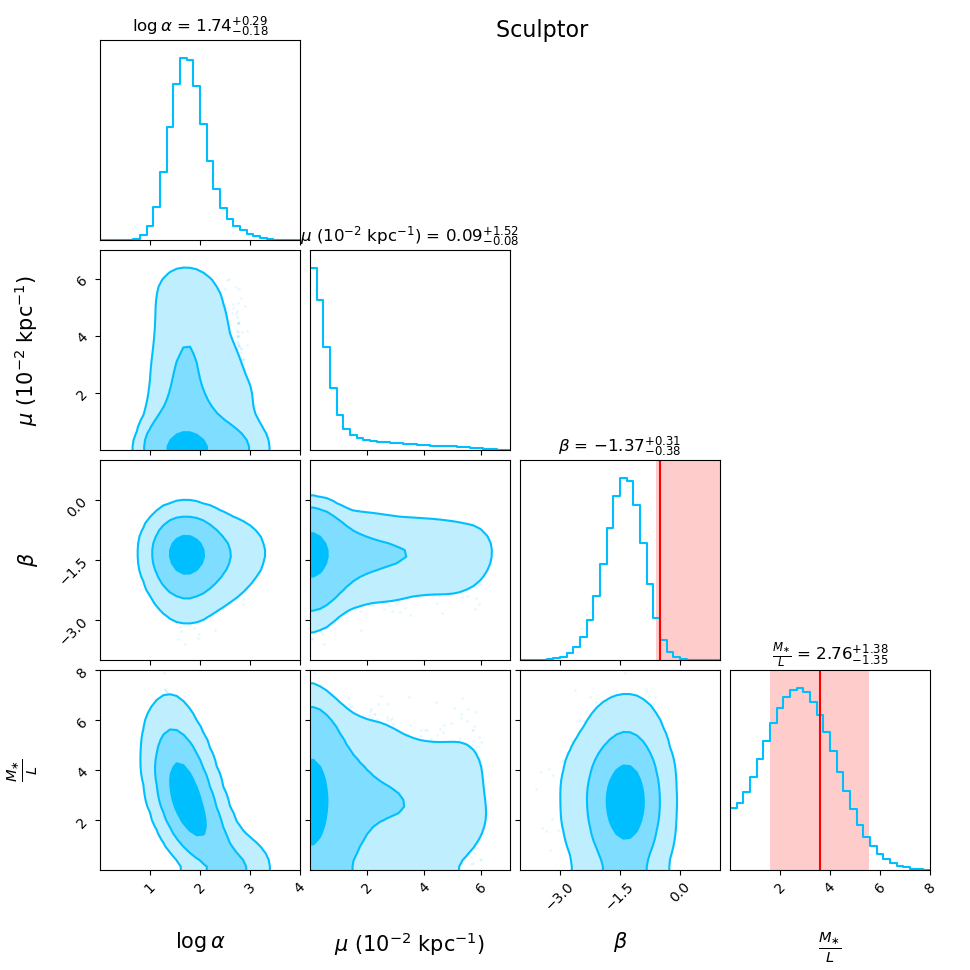}\\
\includegraphics[width=0.99\columnwidth]{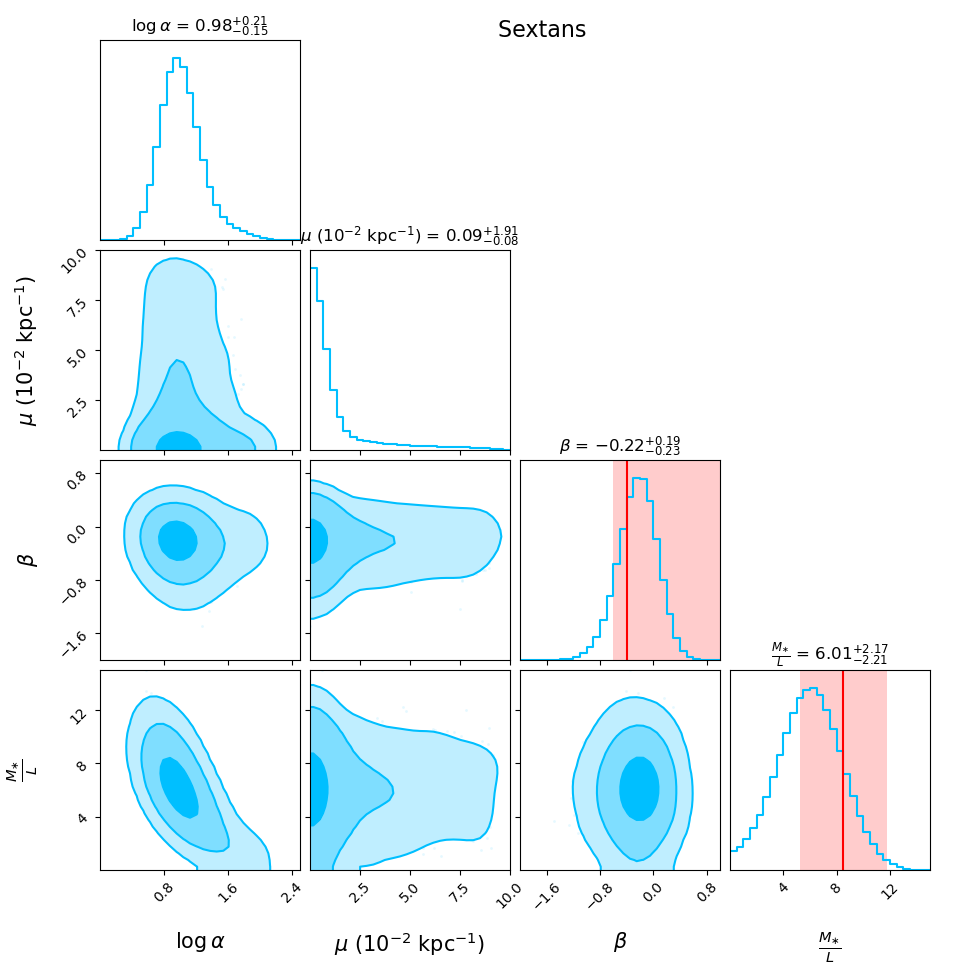}
\includegraphics[width=0.99\columnwidth]{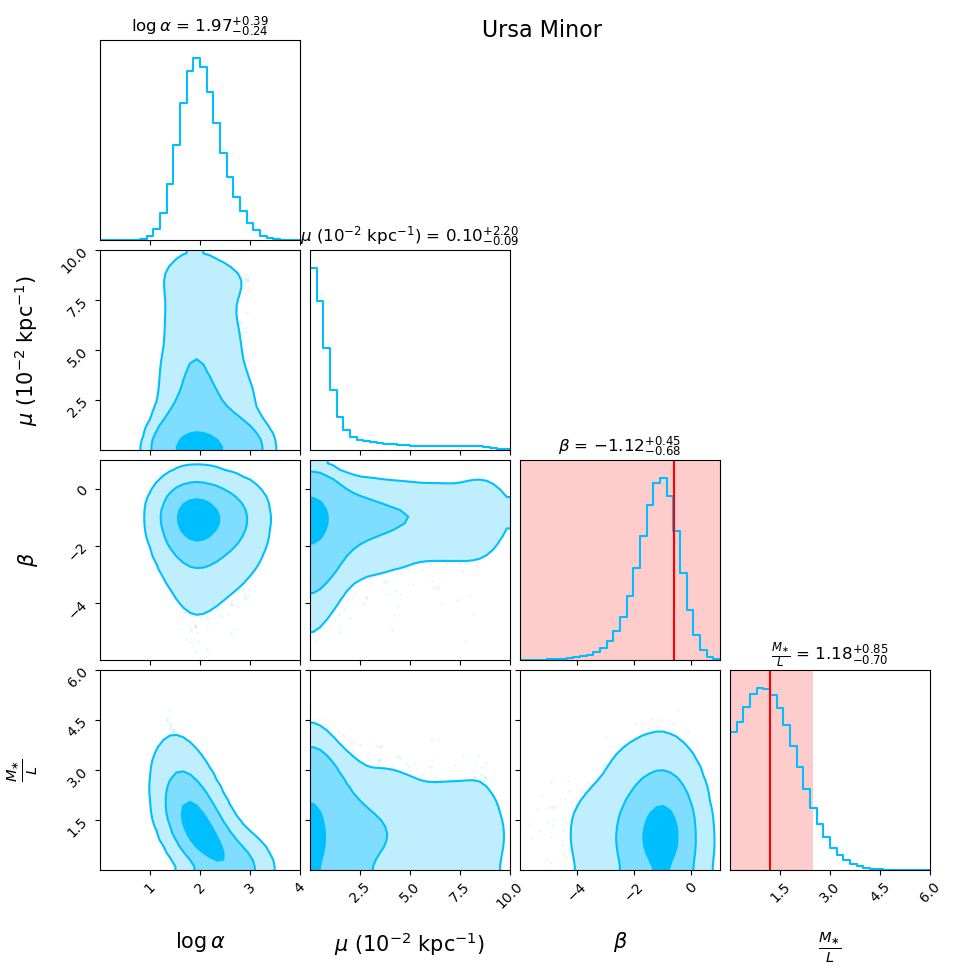}
\end{tabular}
\end{center}
    \caption{{ As in Figure \ref{fig:1}, but particularized for dwarf galaxies Leo II, Sculptor, Sextans, and Ursa Minor.}}\label{fig:2}
\end{figure*}

\begin{figure*}[ht]
\includegraphics[width=2\columnwidth,keepaspectratio]{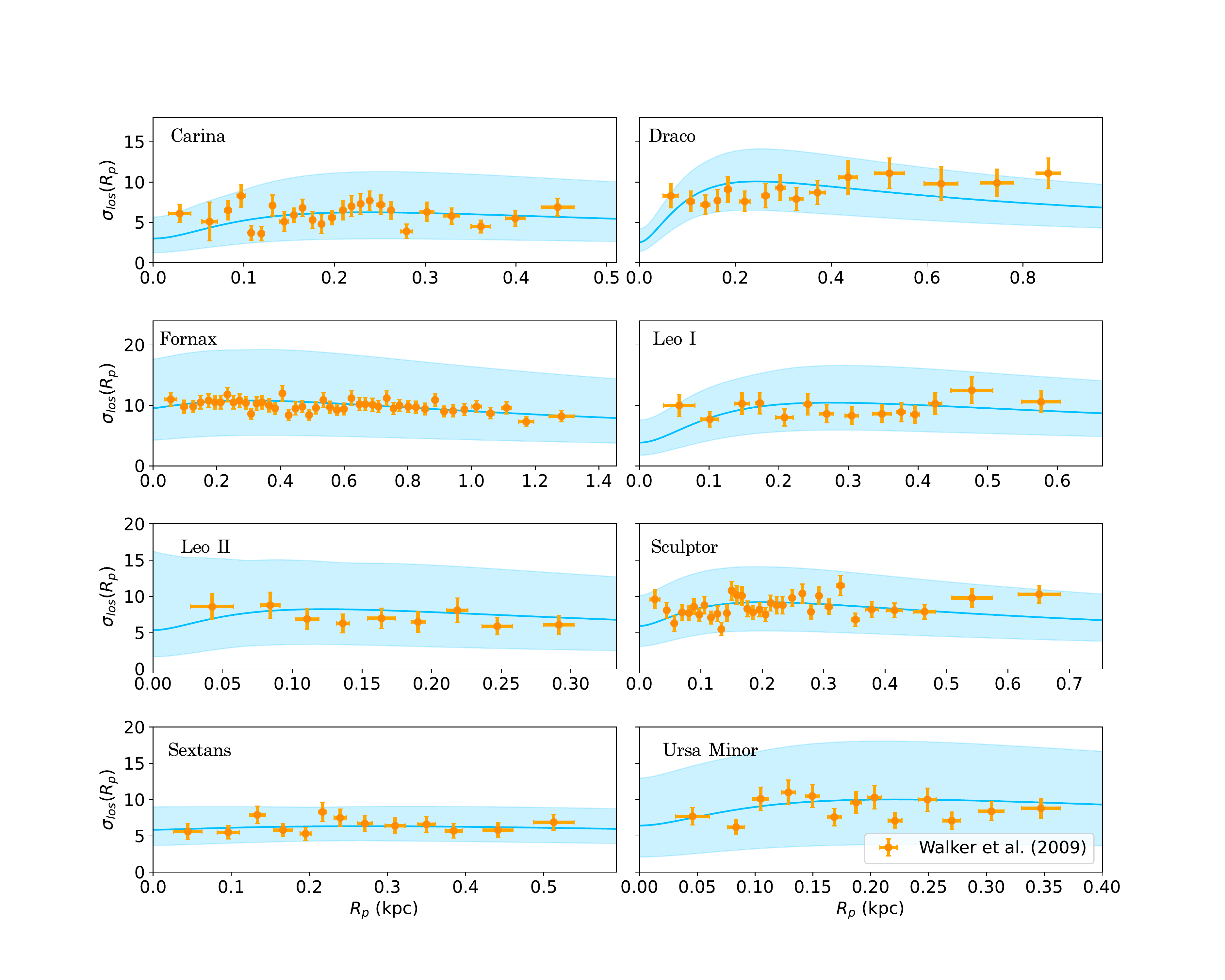}
 \caption{The figure depicts the radial profiles of the line-of-sight velocity dispersions of the eight dSphs  listed in Table \ref{tab:1}. The orange circles with error bars show the measured $\sigma_{\mathrm{los,\, obs}}(R_p)$ from \cite{Walker2009c}. The blue solid lines show the profiles predicted in STVG by adopting the best-fit parameters $\bm{\theta} = \{\log \alpha, \mu, \beta, M_*/L\}$ listed in Table \ref{tab:2}; the blue shaded areas show the  corresponding 68\% confidence interval. }
\label{fig:3}
\end{figure*}

Figures \ref{fig:1} and \ref{fig:2} depict, as blue-shaded areas, the 68\%, 95\%, and 99\% confidence regions with decreasing darkness, respectively. On top of each column, we report the one-dimensional posterior distribution of the corresponding parameter and the median values of the posterior distributions with their 68\% confidence intervals. The red shaded areas correspond to the best-fit values and the $1\sigma$  {uncertainties}  of the velocity anisotropy parameter  reported in \citet{Walker2009d}, and the expected values of $M_*/L$  listed in Table \ref{tab:1}. First, one can note that the stellar mass-to-light ratios agree with the expected values from the stellar population synthesis model. This is somewhat expected as we set a Gaussian prior on $M_*/L$. Second, the anisotropy parameter $\beta$ always reproduces within the 68\% confidence interval the value estimated in the standard CDM model \cite{Walker2009c}. The only exception appears in the dwarf galaxy Sculptor where the agreement is reached only at the 95\% confidence level. In any case, these results point out that the kinematic structure of dwarf galaxies predicted in STVG turns out to be similar to the one expected in the CDM paradigm, {\em i.e.} neither radial nor tangential biases are encountered in STVG with respect to CDM.  Finally, Figure \ref{fig:3} visually shows the effectiveness of the STVG in correctly reproducing the observed line-of-sight velocity dispersion profiles. For each dSph galaxy listed in Table \ref{tab:1}, the orange circles with error bars show the measured $\sigma_{\mathrm{los,\, obs}}(R_{p})$ from \cite{Walker2009c}, the blue solid lines depict the line-of-sight velocity dispersion profiles $\sigma_{\mathrm{los,\, th}}(\bm{\theta},R_{p})$ in STVG gravity predicted by adopting the best-fit parameters $\bm{\theta} = \{\log \alpha, \mu, \beta, M_*/L\}$ listed in Table \ref{tab:2}, and the blue shaded areas show the  corresponding 68\% confidence interval  calculated through Monte Carlo sampling of the one-dimensional posterior distributions shown in Figure \ref{fig:1} and \ref{fig:2}.

\begin{figure*}[ht]
\includegraphics[width=2\columnwidth,keepaspectratio]{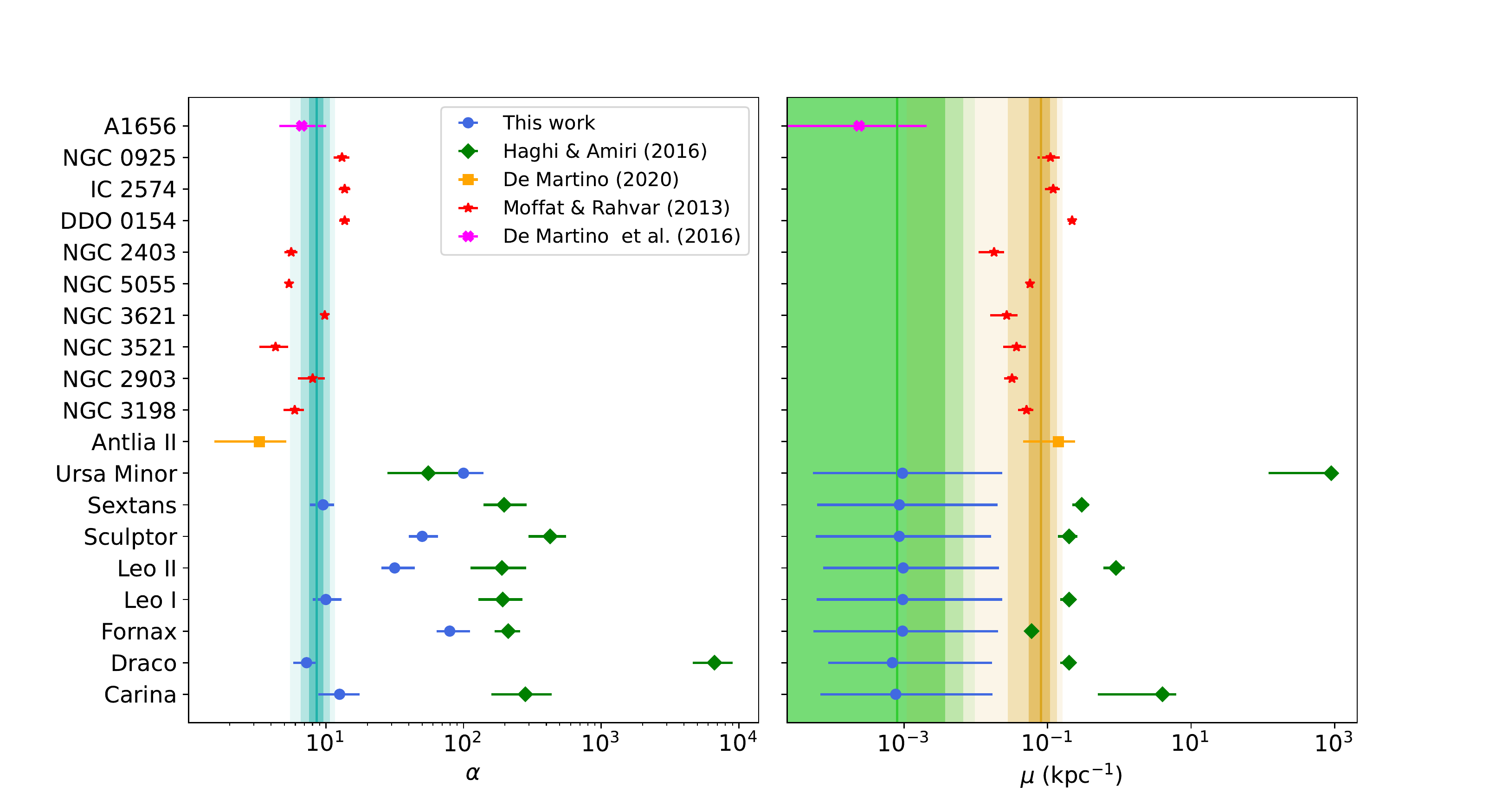}
 \caption{Comparison of the STVG parameters $\alpha$ and $\mu$ obtained in this analysis with previous results, namely: green diamonds reports the estimation of the STVG's parameters obtained using the line-of-sight velocity dispersion of the dSph galaxies by Haghi and Amiri (2016) \cite{Haghi2016}, the gold square depicts the results obtained using the data set of the Antlia II ultra-diffuse galaxy by De Martino (2020) \cite{deMartino2020},  the red stars indicate the estimation of  $\alpha$ and $\mu$ obtained with the rotation curves of spiral galaxies by Moffat and Rahvar (2013) \cite{Moffat2013} and, finally, the magenta point refers to the results obtained using the measured profile of the temperature fluctuations due to the Sunyaev Zel'dovich in the Coma (A1656) cluster by  De Martino and De Laurentis (2016) \cite{DeMartino2016}. In the left panel, the turquoise line represents the average $\alpha$-value of $\overline{\alpha}= 8.60 \pm 1.03$. The average is calculated on  Carina, Draco, Leo I, and Sextans dwarf galaxies, the spiral galaxies used in Moffat and Rahvar (2013) \cite{Moffat2013} and the Coma (A1656) galaxy cluster used  in De Martino and De Laurentis (2016) \cite{DeMartino2016}. On the left panel, the green line represents the average $\mu$-value, namely   $\overline{\mu}= 0.008\pm0.003$ kpc$^{-1}$, calculated using all dSphs and the Coma (A1656) galaxy cluster.
Additionally, the gold line depicts the averaged value $\overline{\mu}= 0.07\pm0.02$ kpc$^{-1}$  obtained by using the estimation of $\mu$ in  Moffat and Rahvar (2013) \cite{Moffat2013} and  De Martino (2020) \cite{deMartino2020}. In both panels, stripes with decreasing darkness depict the 68\%, 95\%, and 99\% confidence regions, respectively. }
\label{fig:4}
\end{figure*}

 Figure \ref{fig:4} illustrates the comparison of the STVG parameters $\alpha$ and $\mu$ obtained in this analysis with previous results obtained using {\em (a)} the line-of-sight velocity dispersion  profiles of the dSph galaxies by Haghi and Amiri (2016) \cite{Haghi2016} and of the Antlia II ultra-diffuse galaxy by De Martino (2020) \cite{deMartino2020},  {\em (b)} the rotation curves of spiral galaxies by Moffat and Rahvar (2013) \cite{Moffat2013} and, finally, {\em (c)} the measured profile of the temperature distortions due to the Sunyaev Zel'dovich effect in the Coma (A1656) cluster by  De Martino and De Laurentis (2016) \cite{DeMartino2016}. First of all, the comparison between our results shown in Table \ref{tab:2} and the estimation of $\alpha$ and $\mu$ obtained  by Haghi and Amiri (2016) \cite{Haghi2016} on the same data sets points out a discrepancy in both parameters ascribable to the different methodology. While we leave the stellar mass-to-light ratio as a free parameter, Haghi and Amiri (2016) fix it to a fiducial value when estimating $\alpha$ and $\mu$. On the other hand, when they leave the stellar mass-to-light ratio  free to vary they fix the value of $\alpha$ and $\mu$ to those obtained by  Moffat and Rahvar (2013) with rotation curves of spiral galaxies \cite{Moffat2013}, and obtain higher values of $M_*/L$ that compensate the missing dark matter halo but do not agree with  the predictions of the stellar population synthesis models \cite{Bell2001}. Interestingly, in our analysis, the value of the parameter $\alpha$ in the cases of Carina, Draco, Leo I, and Sextans agrees within the 68\% confidence interval with all the other estimations found in the literature and shown in  Figure \ref{fig:4}. Therefore, an averaged estimation of the parameter $\alpha$ that would allow STVG to explain the effects ascribable to dark matter on different astrophysical scales would be $\overline{\alpha}= 8.60 \pm 1.03$.  This average value and the corresponding  68\%, 95\%, and 99\% confidence intervals are reported in Figure \ref{fig:4} as the vertical turquoise line and strips (with different darkness levels corresponding to the different confidence intervals).  However, in the case of Fornax, Leo II, Sculptor, and Ursa Minor the estimation of the parameter $\alpha$ shows more than 5$\sigma$ tension with $\overline{\alpha}$. Finally, regarding the parameter $\mu$,  we obtain values in each galaxy  that are compatible with each other, and they also agree with the one obtained in the Coma (A1656) cluster by  De Martino and De Laurentis (2016) \cite{DeMartino2016}. These two data sets give averaged value $\overline{\mu}= 0.008\pm0.003$ kpc$^{-1}$ (reported in Figure \ref{fig:4} as the vertical light green line and strips). However,  the estimated values of the parameter $\mu$ obtained by Moffat and Rahvar (2013) fitting rotation curves of spiral galaxies \cite{Moffat2013} and by  De Martino (2020) \cite{deMartino2020} fitting the velocity dispersion profile of in the Antlia II ultra-diffuse, give an averaged value $\overline{\mu}= 0.07\pm0.02$ kpc$^{-1}$  (depicted in Figure \ref{fig:4} the as the vertical gold line and strips) which is 5$\sigma$ away from the previous one. Nevertheless, this tension in the value of the Yukawa scale length could be ascribable to the different dark matter content required in the CDM model to describe galaxy clusters and dwarf galaxies with respect to spiral galaxies, and therefore figuring out a dependence of the STVG parameters on the mass of the system as argued in \cite{Moffat2013}. Nevertheless, it is worth remarking that no correlation between STVG parameters $\alpha$ and $\mu$ and other observable parameters such as $L_V$, $r_h$, and $M_*/L$ was identified.

\section{Discussion and Conclusions}\label{sec:discussions}

STVG modifies the Einstein-Hilbert action by adding extra massive scalar and vector fields \cite{Moffat2006}. The main aim is to describe the phenomenology of the astrophysical self-gravitating systems without resorting to dark matter. It has been successfully used in several astrophysical scenarios to describe, for instance, the kinematics of stars in galaxies \cite{Moffat2013,Haghi2016,Haghi2019,deMartino2020b}, the mass profile and the Sunyaev-Zel'dovich effect in galaxy clusters \cite{Moffat2014,DeMartino2017}. Although over the years STVG has passed a multitude of probes, some inconsistencies have also arisen. For instance, the analyses based on fitting the line-of-sight velocity dispersion profile of dSph galaxies and of low surface brightness ultra-diffuse galaxies do not provide common values of $\alpha$ and $\mu$  though the masses and luminosities of the galaxies  were comparable to each other \cite{Haghi2016,deMartino2020b}. The reasons can be ascribable to an inappropriate statistical analysis that does not consider all parameters of the model free to vary, or to inappropriate modelling. Indeed, it is well known that dwarf galaxies are not  spherically symmetric but data did not allow for more complicated modelling that would involve the solution of the axisymmetric Jeans equations. Another issue may also be related to the assumption that the anisotropy parameter is taken as a constant while it should depend on the radial position of the star. 

In this work, we investigated whether a more sophisticated statistical analysis of dSph would point out the same inconsistencies of \cite{Haghi2016}. Thus, we have predicted the line-of-sight velocity dispersion profile dSph galaxies listed in Table \ref{tab:1} and fit it to the data taken from \cite{Walker2009c} in order to estimate the value of the two STVG parameters $\alpha$ and $\mu$. Our results turn out to be substantially different from the previous ones that Haghi and Amiri (2016) \cite{Haghi2016} obtained using the same data set. However, our analysis does not fully eliminate the inconsistencies within the dataset  (as shown in the left panel of Figure \ref{fig:4}). The values of $\alpha$ of the dwarf galaxies Carina, Draco, Leo I, and Sextans  agree with each other and with other estimations  coming from rotation curves of spiral galaxies and from galaxy clusters \cite{Moffat2013,DeMartino2017}. The average value for the sample including Carina, Draco, Leo I, and Sextans, all spiral galaxies studied in \cite{Moffat2006} and the Coma (A1656) galaxy cluster is $\overline{\alpha}= 8.60 \pm 1.03$. Finally, the estimation of $\alpha$ in Fornax, Leo II, Sculptor and Ursa Minor show a more than 5$\sigma$ tension with $\overline{\alpha}$. 

This shows an internal inconsistency in the dSph regime. On the other hand, the parameter $\mu$ in each dwarf galaxy turns out to agree with each other and with the estimation coming from the Coma (A1656) cluster  \cite{DeMartino2017} (as shown in the right panel of Figure \ref{fig:4}).  These two data sets give an averaged value $\overline{\mu}= 0.008\pm0.003$ kpc$^{-1}$ (shown as the vertical light green line and strips in Figure \ref{fig:4}). Nevertheless, our results also point out a strong tension with the estimations coming from spiral galaxies. This might be ascribable to different values of the stellar mass-to-light ratio of those self-gravitating systems that would be less dominated by dark matter with respect to dwarf galaxies and galaxy clusters.

In conclusion, our analysis partially solved the inconsistencies highlighted in \cite{Haghi2016} by adopting a more sophisticated statistical treatment. Nevertheless, some issues still remain in the estimation of the parameter $\alpha$. One may expect that the estimation of the STVG parameters might vary on the different astrophysical samples as they could depend on the gravitational mass of the object \cite{Moffat2013}. However, one does not expect STVG parameters to vary within the same class of objects. Further improvements may be obtained with high-precision measurements of the proper motions of the stars belonging the dwarf galaxies that could be available in the future \cite{Theia2017,Malbet2019,Malbet2021}.  These data will allow us  more sophisticated modelling of the line-of-sight velocity dispersion profile where our simplifying assumptions, {\em i.e.} spherical symmetry and $\beta= const.$, might be dropped. 
Consequently, one would expect further improvements in the accuracy of the STVG parameters that could help to fully solve the remaining inconsistencies appearing in dwarf galaxies.

\section*{Acknowledgements}
IDM warmly thanks M. G. Walker for sharing the data set.
IDM acknowledges support from Grant IJCI2018-036198-I  funded by MCIN/AEI/ 10.13039/501100011033 and, as appropriate, by “ESF Investing in your future” or by “European Union NextGenerationEU/PRTR”; and from the  grant PID2021-122938NB-I00  funded by MCIN/AEI/ 10.13039/501100011033 and, as appropriate, by “ERDF A way of making Europe”, by the “European Union” or by the “European Union NextGenerationEU/PRTR”. Finally, IDM acknowledges support from the grant PIC2‐2022‐02 funded by  the University
of Salamanca.

\section*{Data Availability Statement}
Data are publicly available in \cite{Walker2007,Walker2009a,Walker2009b,Walker2009c,Walker2009d}.



\bibliography{refs} 

\begin{thebibliography}{79}%
\makeatletter
\providecommand \@ifxundefined [1]{%
 \@ifx{#1\undefined}
}%
\providecommand \@ifnum [1]{%
 \ifnum #1\expandafter \@firstoftwo
 \else \expandafter \@secondoftwo
 \fi
}%
\providecommand \@ifx [1]{%
 \ifx #1\expandafter \@firstoftwo
 \else \expandafter \@secondoftwo
 \fi
}%
\providecommand \natexlab [1]{#1}%
\providecommand \enquote  [1]{``#1''}%
\providecommand \bibnamefont  [1]{#1}%
\providecommand \bibfnamefont [1]{#1}%
\providecommand \citenamefont [1]{#1}%
\providecommand \href@noop [0]{\@secondoftwo}%
\providecommand \href [0]{\begingroup \@sanitize@url \@href}%
\providecommand \@href[1]{\@@startlink{#1}\@@href}%
\providecommand \@@href[1]{\endgroup#1\@@endlink}%
\providecommand \@sanitize@url [0]{\catcode `\\12\catcode `\$12\catcode
  `\&12\catcode `\#12\catcode `\^12\catcode `\_12\catcode `\%12\relax}%
\providecommand \@@startlink[1]{}%
\providecommand \@@endlink[0]{}%
\providecommand \url  [0]{\begingroup\@sanitize@url \@url }%
\providecommand \@url [1]{\endgroup\@href {#1}{\urlprefix }}%
\providecommand \urlprefix  [0]{URL }%
\providecommand \Eprint [0]{\href }%
\providecommand \doibase [0]{http://dx.doi.org/}%
\providecommand \selectlanguage [0]{\@gobble}%
\providecommand \bibinfo  [0]{\@secondoftwo}%
\providecommand \bibfield  [0]{\@secondoftwo}%
\providecommand \translation [1]{[#1]}%
\providecommand \BibitemOpen [0]{}%
\providecommand \bibitemStop [0]{}%
\providecommand \bibitemNoStop [0]{.\EOS\space}%
\providecommand \EOS [0]{\spacefactor3000\relax}%
\providecommand \BibitemShut  [1]{\csname bibitem#1\endcsname}%
\let\auto@bib@innerbib\@empty
\bibitem [{\citenamefont {{Planck
  Collaboration}}(2020{\natexlab{a}})}]{Planck2020-V}%
  \BibitemOpen
  \bibfield  {author} {\bibinfo {author} {\bibnamefont {{Planck
  Collaboration}}},\ }\href {\doibase 10.1051/0004-6361/201936386} {\bibfield
  {journal} {\bibinfo  {journal} {\aap}\ }\textbf {\bibinfo {volume} {641}},\
  \bibinfo {eid} {A5} (\bibinfo {year} {2020}{\natexlab{a}})},\ \Eprint
  {http://arxiv.org/abs/1907.12875} {arXiv:1907.12875 [astro-ph.CO]}
  \BibitemShut {NoStop}%
\bibitem [{\citenamefont {{Planck
  Collaboration}}(2020{\natexlab{b}})}]{Planck2020-VI}%
  \BibitemOpen
  \bibfield  {author} {\bibinfo {author} {\bibnamefont {{Planck
  Collaboration}}},\ }\href {\doibase 10.1051/0004-6361/201833910} {\bibfield
  {journal} {\bibinfo  {journal} {\aap}\ }\textbf {\bibinfo {volume} {641}},\
  \bibinfo {eid} {A6} (\bibinfo {year} {2020}{\natexlab{b}})},\ \Eprint
  {http://arxiv.org/abs/1807.06209} {arXiv:1807.06209 [astro-ph.CO]}
  \BibitemShut {NoStop}%
\bibitem [{\citenamefont {{Planck
  Collaboration}}(2020{\natexlab{c}})}]{Planck2020-VII}%
  \BibitemOpen
  \bibfield  {author} {\bibinfo {author} {\bibnamefont {{Planck
  Collaboration}}},\ }\href {\doibase 10.1051/0004-6361/201935201} {\bibfield
  {journal} {\bibinfo  {journal} {\aap}\ }\textbf {\bibinfo {volume} {641}},\
  \bibinfo {eid} {A7} (\bibinfo {year} {2020}{\natexlab{c}})},\ \Eprint
  {http://arxiv.org/abs/1906.02552} {arXiv:1906.02552 [astro-ph.CO]}
  \BibitemShut {NoStop}%
\bibitem [{\citenamefont {{Abdalla}}\ and\ \citenamefont {{et
  al.}}(2022)}]{Abdalla2022}%
  \BibitemOpen
  \bibfield  {author} {\bibinfo {author} {\bibfnamefont {E.}~\bibnamefont
  {{Abdalla}}}\ and\ \bibinfo {author} {\bibnamefont {{et al.}}},\ }\href
  {\doibase 10.1016/j.jheap.2022.04.002} {\bibfield  {journal} {\bibinfo
  {journal} {Journal of High Energy Astrophysics}\ }\textbf {\bibinfo {volume}
  {34}},\ \bibinfo {pages} {49} (\bibinfo {year} {2022})},\ \Eprint
  {http://arxiv.org/abs/2203.06142} {arXiv:2203.06142 [astro-ph.CO]}
  \BibitemShut {NoStop}%
\bibitem [{\citenamefont {{Peebles}}(2022)}]{Peebles2022}%
  \BibitemOpen
  \bibfield  {author} {\bibinfo {author} {\bibfnamefont {P.~J.~E.}\
  \bibnamefont {{Peebles}}},\ }\href {\doibase 10.1016/j.aop.2022.169159}
  {\bibfield  {journal} {\bibinfo  {journal} {Annals of Physics}\ }\textbf
  {\bibinfo {volume} {447}},\ \bibinfo {eid} {169159} (\bibinfo {year}
  {2022})},\ \Eprint {http://arxiv.org/abs/2208.05018} {arXiv:2208.05018
  [astro-ph.CO]} \BibitemShut {NoStop}%
\bibitem [{\citenamefont {{Di Valentino}}\ \emph {et~al.}(2021)\citenamefont
  {{Di Valentino}}, \citenamefont {{Mena}}, \citenamefont {{Pan}},
  \citenamefont {{Visinelli}}, \citenamefont {{Yang}}, \citenamefont
  {{Melchiorri}}, \citenamefont {{Mota}}, \citenamefont {{Riess}},\ and\
  \citenamefont {{Silk}}}]{DiValentino2021}%
  \BibitemOpen
  \bibfield  {author} {\bibinfo {author} {\bibfnamefont {E.}~\bibnamefont {{Di
  Valentino}}}, \bibinfo {author} {\bibfnamefont {O.}~\bibnamefont {{Mena}}},
  \bibinfo {author} {\bibfnamefont {S.}~\bibnamefont {{Pan}}}, \bibinfo
  {author} {\bibfnamefont {L.}~\bibnamefont {{Visinelli}}}, \bibinfo {author}
  {\bibfnamefont {W.}~\bibnamefont {{Yang}}}, \bibinfo {author} {\bibfnamefont
  {A.}~\bibnamefont {{Melchiorri}}}, \bibinfo {author} {\bibfnamefont {D.~F.}\
  \bibnamefont {{Mota}}}, \bibinfo {author} {\bibfnamefont {A.~G.}\
  \bibnamefont {{Riess}}}, \ and\ \bibinfo {author} {\bibfnamefont
  {J.}~\bibnamefont {{Silk}}},\ }\href {\doibase 10.1088/1361-6382/ac086d}
  {\bibfield  {journal} {\bibinfo  {journal} {Classical and Quantum Gravity}\
  }\textbf {\bibinfo {volume} {38}},\ \bibinfo {eid} {153001} (\bibinfo {year}
  {2021})},\ \Eprint {http://arxiv.org/abs/2103.01183} {arXiv:2103.01183
  [astro-ph.CO]} \BibitemShut {NoStop}%
\bibitem [{\citenamefont {{Dainotti}}\ \emph {et~al.}(2023)\citenamefont
  {{Dainotti}}, \citenamefont {{De Simone}}, \citenamefont {{Montani}},
  \citenamefont {{Schiavone}},\ and\ \citenamefont
  {{Lambiase}}}]{Dainotti2023}%
  \BibitemOpen
  \bibfield  {author} {\bibinfo {author} {\bibfnamefont {M.}~\bibnamefont
  {{Dainotti}}}, \bibinfo {author} {\bibfnamefont {B.}~\bibnamefont {{De
  Simone}}}, \bibinfo {author} {\bibfnamefont {G.}~\bibnamefont {{Montani}}},
  \bibinfo {author} {\bibfnamefont {T.}~\bibnamefont {{Schiavone}}}, \ and\
  \bibinfo {author} {\bibfnamefont {G.}~\bibnamefont {{Lambiase}}},\ }\href
  {\doibase 10.48550/arXiv.2301.10572} {\bibfield  {journal} {\bibinfo
  {journal} {arXiv e-prints}\ ,\ \bibinfo {eid} {arXiv:2301.10572}} (\bibinfo
  {year} {2023})},\ \Eprint {http://arxiv.org/abs/2301.10572} {arXiv:2301.10572
  [astro-ph.CO]} \BibitemShut {NoStop}%
\bibitem [{\citenamefont {{Califano}}\ \emph {et~al.}(2022)\citenamefont
  {{Califano}}, \citenamefont {{de Martino}}, \citenamefont {{Vernieri}},\ and\
  \citenamefont {{Capozziello}}}]{Califano2022}%
  \BibitemOpen
  \bibfield  {author} {\bibinfo {author} {\bibfnamefont {M.}~\bibnamefont
  {{Califano}}}, \bibinfo {author} {\bibfnamefont {I.}~\bibnamefont {{de
  Martino}}}, \bibinfo {author} {\bibfnamefont {D.}~\bibnamefont {{Vernieri}}},
  \ and\ \bibinfo {author} {\bibfnamefont {S.}~\bibnamefont {{Capozziello}}},\
  }\href {\doibase 10.48550/arXiv.2208.13999} {\bibfield  {journal} {\bibinfo
  {journal} {arXiv e-prints}\ ,\ \bibinfo {eid} {arXiv:2208.13999}} (\bibinfo
  {year} {2022})},\ \Eprint {http://arxiv.org/abs/2208.13999} {arXiv:2208.13999
  [astro-ph.CO]} \BibitemShut {NoStop}%
\bibitem [{\citenamefont {{Boylan-Kolchin}}\ \emph {et~al.}(2011)\citenamefont
  {{Boylan-Kolchin}}, \citenamefont {{Bullock}},\ and\ \citenamefont
  {{Kaplinghat}}}]{Boylan_Kolchin2011}%
  \BibitemOpen
  \bibfield  {author} {\bibinfo {author} {\bibfnamefont {M.}~\bibnamefont
  {{Boylan-Kolchin}}}, \bibinfo {author} {\bibfnamefont {J.~S.}\ \bibnamefont
  {{Bullock}}}, \ and\ \bibinfo {author} {\bibfnamefont {M.}~\bibnamefont
  {{Kaplinghat}}},\ }\href {\doibase 10.1111/j.1745-3933.2011.01074.x}
  {\bibfield  {journal} {\bibinfo  {journal} {\mnras}\ }\textbf {\bibinfo
  {volume} {415}},\ \bibinfo {pages} {L40} (\bibinfo {year} {2011})},\ \Eprint
  {http://arxiv.org/abs/1103.0007} {arXiv:1103.0007 [astro-ph.CO]} \BibitemShut
  {NoStop}%
\bibitem [{\citenamefont {{Bullock}}\ and\ \citenamefont
  {{Boylan-Kolchin}}(2017)}]{Bullock2017}%
  \BibitemOpen
  \bibfield  {author} {\bibinfo {author} {\bibfnamefont {J.~S.}\ \bibnamefont
  {{Bullock}}}\ and\ \bibinfo {author} {\bibfnamefont {M.}~\bibnamefont
  {{Boylan-Kolchin}}},\ }\href {\doibase 10.1146/annurev-astro-091916-055313}
  {\bibfield  {journal} {\bibinfo  {journal} {\araa}\ }\textbf {\bibinfo
  {volume} {55}},\ \bibinfo {pages} {343} (\bibinfo {year} {2017})},\ \Eprint
  {http://arxiv.org/abs/1707.04256} {arXiv:1707.04256 [astro-ph.CO]}
  \BibitemShut {NoStop}%
\bibitem [{\citenamefont {{Del Popolo}}\ and\ \citenamefont {{Le
  Delliou}}(2017)}]{DelPopolo2017}%
  \BibitemOpen
  \bibfield  {author} {\bibinfo {author} {\bibfnamefont {A.}~\bibnamefont {{Del
  Popolo}}}\ and\ \bibinfo {author} {\bibfnamefont {M.}~\bibnamefont {{Le
  Delliou}}},\ }\href {\doibase 10.3390/galaxies5010017} {\bibfield  {journal}
  {\bibinfo  {journal} {Galaxies}\ }\textbf {\bibinfo {volume} {5}},\ \bibinfo
  {pages} {17} (\bibinfo {year} {2017})},\ \Eprint
  {http://arxiv.org/abs/1606.07790} {arXiv:1606.07790 [astro-ph.CO]}
  \BibitemShut {NoStop}%
\bibitem [{\citenamefont {{De Martino}}\ \emph {et~al.}(2020)\citenamefont {{De
  Martino}}, \citenamefont {{Chakrabarty}}, \citenamefont {{Cesare}},
  \citenamefont {{Gallo}}, \citenamefont {{Ostorero}},\ and\ \citenamefont
  {{Diaferio}}}]{deMartino2020}%
  \BibitemOpen
  \bibfield  {author} {\bibinfo {author} {\bibfnamefont {I.}~\bibnamefont {{De
  Martino}}}, \bibinfo {author} {\bibfnamefont {S.~S.}\ \bibnamefont
  {{Chakrabarty}}}, \bibinfo {author} {\bibfnamefont {V.}~\bibnamefont
  {{Cesare}}}, \bibinfo {author} {\bibfnamefont {A.}~\bibnamefont {{Gallo}}},
  \bibinfo {author} {\bibfnamefont {L.}~\bibnamefont {{Ostorero}}}, \ and\
  \bibinfo {author} {\bibfnamefont {A.}~\bibnamefont {{Diaferio}}},\ }\href
  {\doibase 10.3390/universe6080107} {\bibfield  {journal} {\bibinfo  {journal}
  {Universe}\ }\textbf {\bibinfo {volume} {6}},\ \bibinfo {pages} {107}
  (\bibinfo {year} {2020})},\ \Eprint {http://arxiv.org/abs/2007.15539}
  {arXiv:2007.15539 [astro-ph.CO]} \BibitemShut {NoStop}%
\bibitem [{\citenamefont {{Salucci}}\ and\ \citenamefont {{et
  al.}}(2021)}]{Salucci2021}%
  \BibitemOpen
  \bibfield  {author} {\bibinfo {author} {\bibfnamefont {P.}~\bibnamefont
  {{Salucci}}}\ and\ \bibinfo {author} {\bibnamefont {{et al.}}},\ }\href
  {\doibase 10.3389/fphy.2020.603190} {\bibfield  {journal} {\bibinfo
  {journal} {Frontiers in Physics}\ }\textbf {\bibinfo {volume} {8}},\ \bibinfo
  {eid} {579} (\bibinfo {year} {2021})},\ \Eprint
  {http://arxiv.org/abs/2011.09278} {arXiv:2011.09278 [gr-qc]} \BibitemShut
  {NoStop}%
\bibitem [{\citenamefont {{Boldrini}}(2021)}]{Boldrini2021}%
  \BibitemOpen
  \bibfield  {author} {\bibinfo {author} {\bibfnamefont {P.}~\bibnamefont
  {{Boldrini}}},\ }\href {\doibase 10.3390/galaxies10010005} {\bibfield
  {journal} {\bibinfo  {journal} {Galaxies}\ }\textbf {\bibinfo {volume}
  {10}},\ \bibinfo {pages} {5} (\bibinfo {year} {2021})},\ \Eprint
  {http://arxiv.org/abs/2201.01056} {arXiv:2201.01056 [astro-ph.GA]}
  \BibitemShut {NoStop}%
\bibitem [{\citenamefont {{Asencio}}\ \emph {et~al.}(2022)\citenamefont
  {{Asencio}}, \citenamefont {{Banik}}, \citenamefont {{Mieske}}, \citenamefont
  {{Venhola}}, \citenamefont {{Kroupa}},\ and\ \citenamefont
  {{Zhao}}}]{Asencio2022}%
  \BibitemOpen
  \bibfield  {author} {\bibinfo {author} {\bibfnamefont {E.}~\bibnamefont
  {{Asencio}}}, \bibinfo {author} {\bibfnamefont {I.}~\bibnamefont {{Banik}}},
  \bibinfo {author} {\bibfnamefont {S.}~\bibnamefont {{Mieske}}}, \bibinfo
  {author} {\bibfnamefont {A.}~\bibnamefont {{Venhola}}}, \bibinfo {author}
  {\bibfnamefont {P.}~\bibnamefont {{Kroupa}}}, \ and\ \bibinfo {author}
  {\bibfnamefont {H.}~\bibnamefont {{Zhao}}},\ }\href {\doibase
  10.1093/mnras/stac1765} {\bibfield  {journal} {\bibinfo  {journal} {\mnras}\
  }\textbf {\bibinfo {volume} {515}},\ \bibinfo {pages} {2981} (\bibinfo {year}
  {2022})},\ \Eprint {http://arxiv.org/abs/2208.02265} {arXiv:2208.02265
  [astro-ph.GA]} \BibitemShut {NoStop}%
\bibitem [{\citenamefont {{Angus}}(2008)}]{Angus2008}%
  \BibitemOpen
  \bibfield  {author} {\bibinfo {author} {\bibfnamefont {G.~W.}\ \bibnamefont
  {{Angus}}},\ }\href {\doibase 10.1111/j.1365-2966.2008.13351.x} {\bibfield
  {journal} {\bibinfo  {journal} {\mnras}\ }\textbf {\bibinfo {volume} {387}},\
  \bibinfo {pages} {1481} (\bibinfo {year} {2008})},\ \Eprint
  {http://arxiv.org/abs/0804.3812} {arXiv:0804.3812 [astro-ph]} \BibitemShut
  {NoStop}%
\bibitem [{\citenamefont {{Cardone}}\ \emph {et~al.}(2011)\citenamefont
  {{Cardone}}, \citenamefont {{Angus}}, \citenamefont {{Diaferio}},
  \citenamefont {{Tortora}},\ and\ \citenamefont {{Molinaro}}}]{Cardone2011}%
  \BibitemOpen
  \bibfield  {author} {\bibinfo {author} {\bibfnamefont {V.~F.}\ \bibnamefont
  {{Cardone}}}, \bibinfo {author} {\bibfnamefont {G.}~\bibnamefont {{Angus}}},
  \bibinfo {author} {\bibfnamefont {A.}~\bibnamefont {{Diaferio}}}, \bibinfo
  {author} {\bibfnamefont {C.}~\bibnamefont {{Tortora}}}, \ and\ \bibinfo
  {author} {\bibfnamefont {R.}~\bibnamefont {{Molinaro}}},\ }\href {\doibase
  10.1111/j.1365-2966.2010.18081.x} {\bibfield  {journal} {\bibinfo  {journal}
  {\mnras}\ }\textbf {\bibinfo {volume} {412}},\ \bibinfo {pages} {2617}
  (\bibinfo {year} {2011})},\ \Eprint {http://arxiv.org/abs/1011.5741}
  {arXiv:1011.5741 [astro-ph.CO]} \BibitemShut {NoStop}%
\bibitem [{\citenamefont {{Angus}}\ \emph {et~al.}(2012)\citenamefont
  {{Angus}}, \citenamefont {{van der Heyden}},\ and\ \citenamefont
  {{Diaferio}}}]{Angus2012}%
  \BibitemOpen
  \bibfield  {author} {\bibinfo {author} {\bibfnamefont {G.~W.}\ \bibnamefont
  {{Angus}}}, \bibinfo {author} {\bibfnamefont {K.~J.}\ \bibnamefont {{van der
  Heyden}}}, \ and\ \bibinfo {author} {\bibfnamefont {A.}~\bibnamefont
  {{Diaferio}}},\ }\href {\doibase 10.1051/0004-6361/201219189} {\bibfield
  {journal} {\bibinfo  {journal} {\aap}\ }\textbf {\bibinfo {volume} {543}},\
  \bibinfo {eid} {A76} (\bibinfo {year} {2012})},\ \Eprint
  {http://arxiv.org/abs/1303.0995} {arXiv:1303.0995 [astro-ph.GA]} \BibitemShut
  {NoStop}%
\bibitem [{\citenamefont {{Angus}}\ \emph {et~al.}(2014)\citenamefont
  {{Angus}}, \citenamefont {{Gentile}}, \citenamefont {{Diaferio}},
  \citenamefont {{Famaey}},\ and\ \citenamefont {{van der
  Heyden}}}]{Angus2014}%
  \BibitemOpen
  \bibfield  {author} {\bibinfo {author} {\bibfnamefont {G.~W.}\ \bibnamefont
  {{Angus}}}, \bibinfo {author} {\bibfnamefont {G.}~\bibnamefont {{Gentile}}},
  \bibinfo {author} {\bibfnamefont {A.}~\bibnamefont {{Diaferio}}}, \bibinfo
  {author} {\bibfnamefont {B.}~\bibnamefont {{Famaey}}}, \ and\ \bibinfo
  {author} {\bibfnamefont {K.~J.}\ \bibnamefont {{van der Heyden}}},\ }\href
  {\doibase 10.1093/mnras/stu182} {\bibfield  {journal} {\bibinfo  {journal}
  {\mnras}\ }\textbf {\bibinfo {volume} {440}},\ \bibinfo {pages} {746}
  (\bibinfo {year} {2014})},\ \Eprint {http://arxiv.org/abs/1403.4119}
  {arXiv:1403.4119 [astro-ph.GA]} \BibitemShut {NoStop}%
\bibitem [{\citenamefont {{Angus}}\ \emph {et~al.}(2015)\citenamefont
  {{Angus}}, \citenamefont {{Gentile}}, \citenamefont {{Swaters}},
  \citenamefont {{Famaey}}, \citenamefont {{Diaferio}}, \citenamefont
  {{McGaugh}},\ and\ \citenamefont {{Heyden}}}]{Angus2015}%
  \BibitemOpen
  \bibfield  {author} {\bibinfo {author} {\bibfnamefont {G.~W.}\ \bibnamefont
  {{Angus}}}, \bibinfo {author} {\bibfnamefont {G.}~\bibnamefont {{Gentile}}},
  \bibinfo {author} {\bibfnamefont {R.}~\bibnamefont {{Swaters}}}, \bibinfo
  {author} {\bibfnamefont {B.}~\bibnamefont {{Famaey}}}, \bibinfo {author}
  {\bibfnamefont {A.}~\bibnamefont {{Diaferio}}}, \bibinfo {author}
  {\bibfnamefont {S.~S.}\ \bibnamefont {{McGaugh}}}, \ and\ \bibinfo {author}
  {\bibfnamefont {K.~J. v.~d.}\ \bibnamefont {{Heyden}}},\ }\href {\doibase
  10.1093/mnras/stv1132} {\bibfield  {journal} {\bibinfo  {journal} {\mnras}\
  }\textbf {\bibinfo {volume} {451}},\ \bibinfo {pages} {3551} (\bibinfo {year}
  {2015})},\ \Eprint {http://arxiv.org/abs/1505.05522} {arXiv:1505.05522
  [astro-ph.GA]} \BibitemShut {NoStop}%
\bibitem [{\citenamefont {{Chakrabarty}}\ \emph {et~al.}(2022)\citenamefont
  {{Chakrabarty}}, \citenamefont {{Ostorero}}, \citenamefont {{Gallo}},
  \citenamefont {{Ebagezio}},\ and\ \citenamefont
  {{Diaferio}}}]{Chakrabarty2022}%
  \BibitemOpen
  \bibfield  {author} {\bibinfo {author} {\bibfnamefont {S.~S.}\ \bibnamefont
  {{Chakrabarty}}}, \bibinfo {author} {\bibfnamefont {L.}~\bibnamefont
  {{Ostorero}}}, \bibinfo {author} {\bibfnamefont {A.}~\bibnamefont {{Gallo}}},
  \bibinfo {author} {\bibfnamefont {S.}~\bibnamefont {{Ebagezio}}}, \ and\
  \bibinfo {author} {\bibfnamefont {A.}~\bibnamefont {{Diaferio}}},\ }\href
  {\doibase 10.1051/0004-6361/202141136} {\bibfield  {journal} {\bibinfo
  {journal} {\aap}\ }\textbf {\bibinfo {volume} {657}},\ \bibinfo {eid} {A115}
  (\bibinfo {year} {2022})},\ \Eprint {http://arxiv.org/abs/2104.10174}
  {arXiv:2104.10174 [astro-ph.GA]} \BibitemShut {NoStop}%
\bibitem [{\citenamefont {{Famaey}}\ and\ \citenamefont
  {{McGaugh}}(2012)}]{Famaey2012}%
  \BibitemOpen
  \bibfield  {author} {\bibinfo {author} {\bibfnamefont {B.}~\bibnamefont
  {{Famaey}}}\ and\ \bibinfo {author} {\bibfnamefont {S.~S.}\ \bibnamefont
  {{McGaugh}}},\ }\href {\doibase 10.12942/lrr-2012-10} {\bibfield  {journal}
  {\bibinfo  {journal} {Living Reviews in Relativity}\ }\textbf {\bibinfo
  {volume} {15}},\ \bibinfo {eid} {10} (\bibinfo {year} {2012})},\ \Eprint
  {http://arxiv.org/abs/1112.3960} {arXiv:1112.3960 [astro-ph.CO]} \BibitemShut
  {NoStop}%
\bibitem [{\citenamefont {{Banik}}\ and\ \citenamefont
  {{Zhao}}(2022)}]{Banik2022}%
  \BibitemOpen
  \bibfield  {author} {\bibinfo {author} {\bibfnamefont {I.}~\bibnamefont
  {{Banik}}}\ and\ \bibinfo {author} {\bibfnamefont {H.}~\bibnamefont
  {{Zhao}}},\ }\href {\doibase 10.3390/sym14071331} {\bibfield  {journal}
  {\bibinfo  {journal} {Symmetry}\ }\textbf {\bibinfo {volume} {14}},\ \bibinfo
  {pages} {1331} (\bibinfo {year} {2022})},\ \Eprint
  {http://arxiv.org/abs/2110.06936} {arXiv:2110.06936 [astro-ph.CO]}
  \BibitemShut {NoStop}%
\bibitem [{\citenamefont {{Capozziello}}\ \emph {et~al.}(2009)\citenamefont
  {{Capozziello}}, \citenamefont {{de Filippis}},\ and\ \citenamefont
  {{Salzano}}}]{Capozziello2009}%
  \BibitemOpen
  \bibfield  {author} {\bibinfo {author} {\bibfnamefont {S.}~\bibnamefont
  {{Capozziello}}}, \bibinfo {author} {\bibfnamefont {E.}~\bibnamefont {{de
  Filippis}}}, \ and\ \bibinfo {author} {\bibfnamefont {V.}~\bibnamefont
  {{Salzano}}},\ }\href {\doibase 10.1111/j.1365-2966.2008.14382.x} {\bibfield
  {journal} {\bibinfo  {journal} {\mnras}\ }\textbf {\bibinfo {volume} {394}},\
  \bibinfo {pages} {947} (\bibinfo {year} {2009})},\ \Eprint
  {http://arxiv.org/abs/0809.1882} {arXiv:0809.1882 [astro-ph]} \BibitemShut
  {NoStop}%
\bibitem [{\citenamefont {{Capozziello}}\ and\ \citenamefont {{De
  Laurentis}}(2012)}]{Capozziello2012}%
  \BibitemOpen
  \bibfield  {author} {\bibinfo {author} {\bibfnamefont {S.}~\bibnamefont
  {{Capozziello}}}\ and\ \bibinfo {author} {\bibfnamefont {M.}~\bibnamefont
  {{De Laurentis}}},\ }\href {\doibase 10.1002/andp.201200109} {\bibfield
  {journal} {\bibinfo  {journal} {Annalen der Physik}\ }\textbf {\bibinfo
  {volume} {524}},\ \bibinfo {pages} {545} (\bibinfo {year}
  {2012})}\BibitemShut {NoStop}%
\bibitem [{\citenamefont {{Napolitano}}\ \emph {et~al.}(2012)\citenamefont
  {{Napolitano}}, \citenamefont {{Capozziello}}, \citenamefont {{Romanowsky}},
  \citenamefont {{Capaccioli}},\ and\ \citenamefont
  {{Tortora}}}]{Napolitano2012}%
  \BibitemOpen
  \bibfield  {author} {\bibinfo {author} {\bibfnamefont {N.~R.}\ \bibnamefont
  {{Napolitano}}}, \bibinfo {author} {\bibfnamefont {S.}~\bibnamefont
  {{Capozziello}}}, \bibinfo {author} {\bibfnamefont {A.~J.}\ \bibnamefont
  {{Romanowsky}}}, \bibinfo {author} {\bibfnamefont {M.}~\bibnamefont
  {{Capaccioli}}}, \ and\ \bibinfo {author} {\bibfnamefont {C.}~\bibnamefont
  {{Tortora}}},\ }\href {\doibase 10.1088/0004-637X/748/2/87} {\bibfield
  {journal} {\bibinfo  {journal} {\apj}\ }\textbf {\bibinfo {volume} {748}},\
  \bibinfo {eid} {87} (\bibinfo {year} {2012})},\ \Eprint
  {http://arxiv.org/abs/1201.3363} {arXiv:1201.3363 [astro-ph.CO]} \BibitemShut
  {NoStop}%
\bibitem [{\citenamefont {{Stabile}}\ and\ \citenamefont
  {{Capozziello}}(2013)}]{Stabile2013}%
  \BibitemOpen
  \bibfield  {author} {\bibinfo {author} {\bibfnamefont {A.}~\bibnamefont
  {{Stabile}}}\ and\ \bibinfo {author} {\bibfnamefont {S.}~\bibnamefont
  {{Capozziello}}},\ }\href {\doibase 10.1103/PhysRevD.87.064002} {\bibfield
  {journal} {\bibinfo  {journal} {\prd}\ }\textbf {\bibinfo {volume} {87}},\
  \bibinfo {eid} {064002} (\bibinfo {year} {2013})},\ \Eprint
  {http://arxiv.org/abs/1302.1760} {arXiv:1302.1760 [gr-qc]} \BibitemShut
  {NoStop}%
\bibitem [{\citenamefont {{De Martino}}\ \emph {et~al.}(2014)\citenamefont {{De
  Martino}}, \citenamefont {{De Laurentis}}, \citenamefont
  {{Atrio-Barandela}},\ and\ \citenamefont {{Capozziello}}}]{DeMartino2014}%
  \BibitemOpen
  \bibfield  {author} {\bibinfo {author} {\bibfnamefont {I.}~\bibnamefont {{De
  Martino}}}, \bibinfo {author} {\bibfnamefont {M.}~\bibnamefont {{De
  Laurentis}}}, \bibinfo {author} {\bibfnamefont {F.}~\bibnamefont
  {{Atrio-Barandela}}}, \ and\ \bibinfo {author} {\bibfnamefont
  {S.}~\bibnamefont {{Capozziello}}},\ }\href {\doibase 10.1093/mnras/stu903}
  {\bibfield  {journal} {\bibinfo  {journal} {\mnras}\ }\textbf {\bibinfo
  {volume} {442}},\ \bibinfo {pages} {921} (\bibinfo {year} {2014})},\ \Eprint
  {http://arxiv.org/abs/1310.0693} {arXiv:1310.0693 [astro-ph.CO]} \BibitemShut
  {NoStop}%
\bibitem [{\citenamefont {{De Martino}}(2016)}]{DeMartino2016}%
  \BibitemOpen
  \bibfield  {author} {\bibinfo {author} {\bibfnamefont {I.}~\bibnamefont {{De
  Martino}}},\ }\href {\doibase 10.1103/PhysRevD.93.124043} {\bibfield
  {journal} {\bibinfo  {journal} {\prd}\ }\textbf {\bibinfo {volume} {93}},\
  \bibinfo {eid} {124043} (\bibinfo {year} {2016})},\ \Eprint
  {http://arxiv.org/abs/1605.08223} {arXiv:1605.08223 [astro-ph.CO]}
  \BibitemShut {NoStop}%
\bibitem [{\citenamefont {{de Martino}}\ \emph {et~al.}(2023)\citenamefont {{de
  Martino}}, \citenamefont {{Diaferio}},\ and\ \citenamefont
  {{Ostorero}}}]{deMartino2023}%
  \BibitemOpen
  \bibfield  {author} {\bibinfo {author} {\bibfnamefont {I.}~\bibnamefont {{de
  Martino}}}, \bibinfo {author} {\bibfnamefont {A.}~\bibnamefont {{Diaferio}}},
  \ and\ \bibinfo {author} {\bibfnamefont {L.}~\bibnamefont {{Ostorero}}},\
  }\href {\doibase 10.1093/mnras/stad010} {\bibfield  {journal} {\bibinfo
  {journal} {\mnras}\ }\textbf {\bibinfo {volume} {519}},\ \bibinfo {pages}
  {4424} (\bibinfo {year} {2023})},\ \Eprint {http://arxiv.org/abs/2210.02306}
  {arXiv:2210.02306 [astro-ph.GA]} \BibitemShut {NoStop}%
\bibitem [{\citenamefont {{Laudato}}\ and\ \citenamefont
  {{Salzano}}(2022{\natexlab{a}})}]{Laudato2022a}%
  \BibitemOpen
  \bibfield  {author} {\bibinfo {author} {\bibfnamefont {E.}~\bibnamefont
  {{Laudato}}}\ and\ \bibinfo {author} {\bibfnamefont {V.}~\bibnamefont
  {{Salzano}}},\ }\href {\doibase 10.1140/epjc/s10052-022-10901-0} {\bibfield
  {journal} {\bibinfo  {journal} {European Physical Journal C}\ }\textbf
  {\bibinfo {volume} {82}},\ \bibinfo {eid} {935} (\bibinfo {year}
  {2022}{\natexlab{a}})},\ \Eprint {http://arxiv.org/abs/2206.06284}
  {arXiv:2206.06284 [gr-qc]} \BibitemShut {NoStop}%
\bibitem [{\citenamefont {{Laudato}}\ and\ \citenamefont
  {{Salzano}}(2022{\natexlab{b}})}]{Laudato2022}%
  \BibitemOpen
  \bibfield  {author} {\bibinfo {author} {\bibfnamefont {E.}~\bibnamefont
  {{Laudato}}}\ and\ \bibinfo {author} {\bibfnamefont {V.}~\bibnamefont
  {{Salzano}}},\ }\href {\doibase 10.48550/arXiv.2211.08839} {\bibfield
  {journal} {\bibinfo  {journal} {arXiv e-prints}\ ,\ \bibinfo {eid}
  {arXiv:2211.08839}} (\bibinfo {year} {2022}{\natexlab{b}})},\ \Eprint
  {http://arxiv.org/abs/2211.08839} {arXiv:2211.08839 [gr-qc]} \BibitemShut
  {NoStop}%
\bibitem [{\citenamefont {{Cesare}}\ \emph {et~al.}(2020)\citenamefont
  {{Cesare}}, \citenamefont {{Diaferio}}, \citenamefont {{Matsakos}},\ and\
  \citenamefont {{Angus}}}]{Cesare2020}%
  \BibitemOpen
  \bibfield  {author} {\bibinfo {author} {\bibfnamefont {V.}~\bibnamefont
  {{Cesare}}}, \bibinfo {author} {\bibfnamefont {A.}~\bibnamefont
  {{Diaferio}}}, \bibinfo {author} {\bibfnamefont {T.}~\bibnamefont
  {{Matsakos}}}, \ and\ \bibinfo {author} {\bibfnamefont {G.}~\bibnamefont
  {{Angus}}},\ }\href {\doibase 10.1051/0004-6361/201935950} {\bibfield
  {journal} {\bibinfo  {journal} {\aap}\ }\textbf {\bibinfo {volume} {637}},\
  \bibinfo {eid} {A70} (\bibinfo {year} {2020})},\ \Eprint
  {http://arxiv.org/abs/2003.07377} {arXiv:2003.07377 [astro-ph.GA]}
  \BibitemShut {NoStop}%
\bibitem [{\citenamefont {{Cesare}}\ \emph {et~al.}(2022)\citenamefont
  {{Cesare}}, \citenamefont {{Diaferio}},\ and\ \citenamefont
  {{Matsakos}}}]{Cesare2022}%
  \BibitemOpen
  \bibfield  {author} {\bibinfo {author} {\bibfnamefont {V.}~\bibnamefont
  {{Cesare}}}, \bibinfo {author} {\bibfnamefont {A.}~\bibnamefont
  {{Diaferio}}}, \ and\ \bibinfo {author} {\bibfnamefont {T.}~\bibnamefont
  {{Matsakos}}},\ }\href {\doibase 10.1051/0004-6361/202140651} {\bibfield
  {journal} {\bibinfo  {journal} {\aap}\ }\textbf {\bibinfo {volume} {657}},\
  \bibinfo {eid} {A133} (\bibinfo {year} {2022})},\ \Eprint
  {http://arxiv.org/abs/2102.12499} {arXiv:2102.12499 [astro-ph.GA]}
  \BibitemShut {NoStop}%
\bibitem [{\citenamefont {{Pizzuti}}\ \emph {et~al.}(2022)\citenamefont
  {{Pizzuti}}, \citenamefont {{Saltas}}, \citenamefont {{Umetsu}},\ and\
  \citenamefont {{Sartoris}}}]{Pizzuti2022a}%
  \BibitemOpen
  \bibfield  {author} {\bibinfo {author} {\bibfnamefont {L.}~\bibnamefont
  {{Pizzuti}}}, \bibinfo {author} {\bibfnamefont {I.~D.}\ \bibnamefont
  {{Saltas}}}, \bibinfo {author} {\bibfnamefont {K.}~\bibnamefont {{Umetsu}}},
  \ and\ \bibinfo {author} {\bibfnamefont {B.}~\bibnamefont {{Sartoris}}},\
  }\href {\doibase 10.1093/mnras/stac746} {\bibfield  {journal} {\bibinfo
  {journal} {\mnras}\ }\textbf {\bibinfo {volume} {512}},\ \bibinfo {pages}
  {4280} (\bibinfo {year} {2022})},\ \Eprint {http://arxiv.org/abs/2112.12139}
  {arXiv:2112.12139 [astro-ph.CO]} \BibitemShut {NoStop}%
\bibitem [{\citenamefont {{Pizzuti}}(2022)}]{Pizzuti2022b}%
  \BibitemOpen
  \bibfield  {author} {\bibinfo {author} {\bibfnamefont {L.}~\bibnamefont
  {{Pizzuti}}},\ }\href {\doibase 10.3390/universe8030157} {\bibfield
  {journal} {\bibinfo  {journal} {Universe}\ }\textbf {\bibinfo {volume} {8}},\
  \bibinfo {pages} {157} (\bibinfo {year} {2022})},\ \Eprint
  {http://arxiv.org/abs/2204.04432} {arXiv:2204.04432 [astro-ph.CO]}
  \BibitemShut {NoStop}%
\bibitem [{\citenamefont {{Moffat}}(2006)}]{Moffat2006}%
  \BibitemOpen
  \bibfield  {author} {\bibinfo {author} {\bibfnamefont {J.~W.}\ \bibnamefont
  {{Moffat}}},\ }\href {\doibase 10.1088/1475-7516/2006/03/004} {\bibfield
  {journal} {\bibinfo  {journal} {\jcap}\ }\textbf {\bibinfo {volume} {2006}},\
  \bibinfo {eid} {004} (\bibinfo {year} {2006})},\ \Eprint
  {http://arxiv.org/abs/gr-qc/0506021} {arXiv:gr-qc/0506021 [gr-qc]}
  \BibitemShut {NoStop}%
\bibitem [{\citenamefont {{Moffat}}\ and\ \citenamefont
  {{Rahvar}}(2013)}]{Moffat2013a}%
  \BibitemOpen
  \bibfield  {author} {\bibinfo {author} {\bibfnamefont {J.~W.}\ \bibnamefont
  {{Moffat}}}\ and\ \bibinfo {author} {\bibfnamefont {S.}~\bibnamefont
  {{Rahvar}}},\ }\href {\doibase 10.1093/mnras/stt1670} {\bibfield  {journal}
  {\bibinfo  {journal} {\mnras}\ }\textbf {\bibinfo {volume} {436}},\ \bibinfo
  {pages} {1439} (\bibinfo {year} {2013})},\ \Eprint
  {http://arxiv.org/abs/1306.6383} {arXiv:1306.6383 [astro-ph.GA]} \BibitemShut
  {NoStop}%
\bibitem [{\citenamefont {{Moffat}}\ and\ \citenamefont
  {{Rahvar}}(2014)}]{Moffat2014}%
  \BibitemOpen
  \bibfield  {author} {\bibinfo {author} {\bibfnamefont {J.~W.}\ \bibnamefont
  {{Moffat}}}\ and\ \bibinfo {author} {\bibfnamefont {S.}~\bibnamefont
  {{Rahvar}}},\ }\href {\doibase 10.1093/mnras/stu855} {\bibfield  {journal}
  {\bibinfo  {journal} {\mnras}\ }\textbf {\bibinfo {volume} {441}},\ \bibinfo
  {pages} {3724} (\bibinfo {year} {2014})},\ \Eprint
  {http://arxiv.org/abs/1309.5077} {arXiv:1309.5077 [astro-ph.CO]} \BibitemShut
  {NoStop}%
\bibitem [{\citenamefont {{Della Monica}}\ \emph {et~al.}(2022)\citenamefont
  {{Della Monica}}, \citenamefont {{de Martino}},\ and\ \citenamefont {{de
  Laurentis}}}]{DellaMonica2022}%
  \BibitemOpen
  \bibfield  {author} {\bibinfo {author} {\bibfnamefont {R.}~\bibnamefont
  {{Della Monica}}}, \bibinfo {author} {\bibfnamefont {I.}~\bibnamefont {{de
  Martino}}}, \ and\ \bibinfo {author} {\bibfnamefont {M.}~\bibnamefont {{de
  Laurentis}}},\ }\href {\doibase 10.1093/mnras/stab3727} {\bibfield  {journal}
  {\bibinfo  {journal} {\mnras}\ }\textbf {\bibinfo {volume} {510}},\ \bibinfo
  {pages} {4757} (\bibinfo {year} {2022})},\ \Eprint
  {http://arxiv.org/abs/2105.12687} {arXiv:2105.12687 [gr-qc]} \BibitemShut
  {NoStop}%
\bibitem [{\citenamefont {{Della Monica}}\ \emph {et~al.}(2023)\citenamefont
  {{Della Monica}}, \citenamefont {{de Martino}},\ and\ \citenamefont {{de
  Laurentis}}}]{DellaMonica2023}%
  \BibitemOpen
  \bibfield  {author} {\bibinfo {author} {\bibfnamefont {R.}~\bibnamefont
  {{Della Monica}}}, \bibinfo {author} {\bibfnamefont {I.}~\bibnamefont {{de
  Martino}}}, \ and\ \bibinfo {author} {\bibfnamefont {M.}~\bibnamefont {{de
  Laurentis}}},\ }\href {\doibase 10.1093/mnras/stad579} {\bibfield  {journal}
  {\bibinfo  {journal} {\mnras}\ } (\bibinfo {year} {2023}),\
  10.1093/mnras/stad579},\ \Eprint {http://arxiv.org/abs/2302.12296}
  {arXiv:2302.12296 [gr-qc]} \BibitemShut {NoStop}%
\bibitem [{\citenamefont {{Ghafourian}}\ and\ \citenamefont
  {{Roshan}}(2017)}]{Ghafourian2017}%
  \BibitemOpen
  \bibfield  {author} {\bibinfo {author} {\bibfnamefont {N.}~\bibnamefont
  {{Ghafourian}}}\ and\ \bibinfo {author} {\bibfnamefont {M.}~\bibnamefont
  {{Roshan}}},\ }\href {\doibase 10.1093/mnras/stx661} {\bibfield  {journal}
  {\bibinfo  {journal} {\mnras}\ }\textbf {\bibinfo {volume} {468}},\ \bibinfo
  {pages} {4450} (\bibinfo {year} {2017})}\BibitemShut {NoStop}%
\bibitem [{\citenamefont {{Ghafourian}}\ and\ \citenamefont
  {{Roshan}}(2020)}]{Ghafourian2020}%
  \BibitemOpen
  \bibfield  {author} {\bibinfo {author} {\bibfnamefont {N.}~\bibnamefont
  {{Ghafourian}}}\ and\ \bibinfo {author} {\bibfnamefont {M.}~\bibnamefont
  {{Roshan}}},\ }in\ \href {\doibase 10.1017/S1743921319008810} {\emph
  {\bibinfo {booktitle} {Galactic Dynamics in the Era of Large Surveys}}},\
  Vol.\ \bibinfo {volume} {353},\ \bibinfo {editor} {edited by\ \bibinfo
  {editor} {\bibfnamefont {M.}~\bibnamefont {{Valluri}}}\ and\ \bibinfo
  {editor} {\bibfnamefont {J.~A.}\ \bibnamefont {{Sellwood}}}}\ (\bibinfo
  {year} {2020})\ pp.\ \bibinfo {pages} {152--153}\BibitemShut {NoStop}%
\bibitem [{\citenamefont {{Moffat}}\ and\ \citenamefont
  {{Toth}}(2013)}]{Moffat2013}%
  \BibitemOpen
  \bibfield  {author} {\bibinfo {author} {\bibfnamefont {J.}~\bibnamefont
  {{Moffat}}}\ and\ \bibinfo {author} {\bibfnamefont {V.}~\bibnamefont
  {{Toth}}},\ }\href {\doibase 10.3390/galaxies1010065} {\bibfield  {journal}
  {\bibinfo  {journal} {Galaxies}\ }\textbf {\bibinfo {volume} {1}},\ \bibinfo
  {pages} {65} (\bibinfo {year} {2013})}\BibitemShut {NoStop}%
\bibitem [{\citenamefont {{Moffat}}(2016)}]{Moffat2016}%
  \BibitemOpen
  \bibfield  {author} {\bibinfo {author} {\bibfnamefont {J.~W.}\ \bibnamefont
  {{Moffat}}},\ }\href {\doibase 10.1016/j.physletb.2016.10.082} {\bibfield
  {journal} {\bibinfo  {journal} {Physics Letters B}\ }\textbf {\bibinfo
  {volume} {763}},\ \bibinfo {pages} {427} (\bibinfo {year} {2016})},\ \Eprint
  {http://arxiv.org/abs/1603.05225} {arXiv:1603.05225 [gr-qc]} \BibitemShut
  {NoStop}%
\bibitem [{\citenamefont {{De Martino}}\ and\ \citenamefont {{De
  Laurentis}}(2017)}]{DeMartino2017}%
  \BibitemOpen
  \bibfield  {author} {\bibinfo {author} {\bibfnamefont {I.}~\bibnamefont {{De
  Martino}}}\ and\ \bibinfo {author} {\bibfnamefont {M.}~\bibnamefont {{De
  Laurentis}}},\ }\href {\doibase 10.1016/j.physletb.2017.05.012} {\bibfield
  {journal} {\bibinfo  {journal} {Physics Letters B}\ }\textbf {\bibinfo
  {volume} {770}},\ \bibinfo {pages} {440} (\bibinfo {year} {2017})},\ \Eprint
  {http://arxiv.org/abs/1705.02366} {arXiv:1705.02366 [astro-ph.CO]}
  \BibitemShut {NoStop}%
\bibitem [{\citenamefont {{Haghi}}\ and\ \citenamefont
  {{Amiri}}(2016)}]{Haghi2016}%
  \BibitemOpen
  \bibfield  {author} {\bibinfo {author} {\bibfnamefont {H.}~\bibnamefont
  {{Haghi}}}\ and\ \bibinfo {author} {\bibfnamefont {V.}~\bibnamefont
  {{Amiri}}},\ }\href {\doibase 10.1093/mnras/stw2140} {\bibfield  {journal}
  {\bibinfo  {journal} {\mnras}\ }\textbf {\bibinfo {volume} {463}},\ \bibinfo
  {pages} {1944} (\bibinfo {year} {2016})},\ \Eprint
  {http://arxiv.org/abs/1609.03238} {arXiv:1609.03238 [astro-ph.GA]}
  \BibitemShut {NoStop}%
\bibitem [{\citenamefont {{de Martino}}(2020)}]{deMartino2020b}%
  \BibitemOpen
  \bibfield  {author} {\bibinfo {author} {\bibfnamefont {I.}~\bibnamefont {{de
  Martino}}},\ }\href {\doibase 10.1093/mnras/staa460} {\bibfield  {journal}
  {\bibinfo  {journal} {\mnras}\ }\textbf {\bibinfo {volume} {493}},\ \bibinfo
  {pages} {2373} (\bibinfo {year} {2020})},\ \Eprint
  {http://arxiv.org/abs/2002.05161} {arXiv:2002.05161 [gr-qc]} \BibitemShut
  {NoStop}%
\bibitem [{\citenamefont {{Bell}}\ and\ \citenamefont {{de
  Jong}}(2001)}]{Bell2001}%
  \BibitemOpen
  \bibfield  {author} {\bibinfo {author} {\bibfnamefont {E.~F.}\ \bibnamefont
  {{Bell}}}\ and\ \bibinfo {author} {\bibfnamefont {R.~S.}\ \bibnamefont {{de
  Jong}}},\ }\href {\doibase 10.1086/319728} {\bibfield  {journal} {\bibinfo
  {journal} {\apj}\ }\textbf {\bibinfo {volume} {550}},\ \bibinfo {pages} {212}
  (\bibinfo {year} {2001})},\ \Eprint {http://arxiv.org/abs/astro-ph/0011493}
  {arXiv:astro-ph/0011493 [astro-ph]} \BibitemShut {NoStop}%
\bibitem [{\citenamefont {{Zhoolideh Haghighi}}\ and\ \citenamefont
  {{Rahvar}}(2017)}]{Haghighi2017}%
  \BibitemOpen
  \bibfield  {author} {\bibinfo {author} {\bibfnamefont {M.~H.}\ \bibnamefont
  {{Zhoolideh Haghighi}}}\ and\ \bibinfo {author} {\bibfnamefont
  {S.}~\bibnamefont {{Rahvar}}},\ }\href {\doibase 10.1093/mnras/stx692}
  {\bibfield  {journal} {\bibinfo  {journal} {\mnras}\ }\textbf {\bibinfo
  {volume} {468}},\ \bibinfo {pages} {4048} (\bibinfo {year} {2017})},\ \Eprint
  {http://arxiv.org/abs/1609.07851} {arXiv:1609.07851 [astro-ph.GA]}
  \BibitemShut {NoStop}%
\bibitem [{\citenamefont {{Haghi}}\ \emph {et~al.}(2019)\citenamefont
  {{Haghi}}, \citenamefont {{Amiri}}, \citenamefont {{Hasani Zonoozi}},
  \citenamefont {{Banik}}, \citenamefont {{Kroupa}},\ and\ \citenamefont
  {{Haslbauer}}}]{Haghi2019}%
  \BibitemOpen
  \bibfield  {author} {\bibinfo {author} {\bibfnamefont {H.}~\bibnamefont
  {{Haghi}}}, \bibinfo {author} {\bibfnamefont {V.}~\bibnamefont {{Amiri}}},
  \bibinfo {author} {\bibfnamefont {A.}~\bibnamefont {{Hasani Zonoozi}}},
  \bibinfo {author} {\bibfnamefont {I.}~\bibnamefont {{Banik}}}, \bibinfo
  {author} {\bibfnamefont {P.}~\bibnamefont {{Kroupa}}}, \ and\ \bibinfo
  {author} {\bibfnamefont {M.}~\bibnamefont {{Haslbauer}}},\ }\href {\doibase
  10.3847/2041-8213/ab4517} {\bibfield  {journal} {\bibinfo  {journal} {\apjl}\
  }\textbf {\bibinfo {volume} {884}},\ \bibinfo {eid} {L25} (\bibinfo {year}
  {2019})},\ \Eprint {http://arxiv.org/abs/1909.07978} {arXiv:1909.07978
  [astro-ph.GA]} \BibitemShut {NoStop}%
\bibitem [{\citenamefont {{Negrelli}}\ \emph {et~al.}(2018)\citenamefont
  {{Negrelli}}, \citenamefont {{Benito}}, \citenamefont {{Landau}},
  \citenamefont {{Iocco}},\ and\ \citenamefont {{Kraiselburd}}}]{Negrelli2018}%
  \BibitemOpen
  \bibfield  {author} {\bibinfo {author} {\bibfnamefont {C.}~\bibnamefont
  {{Negrelli}}}, \bibinfo {author} {\bibfnamefont {M.}~\bibnamefont
  {{Benito}}}, \bibinfo {author} {\bibfnamefont {S.}~\bibnamefont {{Landau}}},
  \bibinfo {author} {\bibfnamefont {F.}~\bibnamefont {{Iocco}}}, \ and\
  \bibinfo {author} {\bibfnamefont {L.}~\bibnamefont {{Kraiselburd}}},\ }\href
  {\doibase 10.1103/PhysRevD.98.104061} {\bibfield  {journal} {\bibinfo
  {journal} {\prd}\ }\textbf {\bibinfo {volume} {98}},\ \bibinfo {eid} {104061}
  (\bibinfo {year} {2018})},\ \Eprint {http://arxiv.org/abs/1810.07200}
  {arXiv:1810.07200 [astro-ph.GA]} \BibitemShut {NoStop}%
\bibitem [{\citenamefont {{Banerjee}}\ \emph {et~al.}(2017)\citenamefont
  {{Banerjee}}, \citenamefont {{Shankar}},\ and\ \citenamefont
  {{Singh}}}]{Banerjee2017}%
  \BibitemOpen
  \bibfield  {author} {\bibinfo {author} {\bibfnamefont {S.}~\bibnamefont
  {{Banerjee}}}, \bibinfo {author} {\bibfnamefont {S.}~\bibnamefont
  {{Shankar}}}, \ and\ \bibinfo {author} {\bibfnamefont {T.~P.}\ \bibnamefont
  {{Singh}}},\ }\href {\doibase 10.1088/1475-7516/2017/10/004} {\bibfield
  {journal} {\bibinfo  {journal} {\jcap}\ }\textbf {\bibinfo {volume} {2017}},\
  \bibinfo {eid} {004} (\bibinfo {year} {2017})},\ \Eprint
  {http://arxiv.org/abs/1705.01048} {arXiv:1705.01048 [gr-qc]} \BibitemShut
  {NoStop}%
\bibitem [{\citenamefont {{{\L}okas}}\ and\ \citenamefont
  {{Mamon}}(2003)}]{Lokas2003}%
  \BibitemOpen
  \bibfield  {author} {\bibinfo {author} {\bibfnamefont {E.~L.}\ \bibnamefont
  {{{\L}okas}}}\ and\ \bibinfo {author} {\bibfnamefont {G.~A.}\ \bibnamefont
  {{Mamon}}},\ }\href {\doibase 10.1046/j.1365-8711.2003.06684.x} {\bibfield
  {journal} {\bibinfo  {journal} {\mnras}\ }\textbf {\bibinfo {volume} {343}},\
  \bibinfo {pages} {401} (\bibinfo {year} {2003})},\ \Eprint
  {http://arxiv.org/abs/astro-ph/0302461} {arXiv:astro-ph/0302461 [astro-ph]}
  \BibitemShut {NoStop}%
\bibitem [{\citenamefont {{Mamon}}\ and\ \citenamefont
  {{{\L}okas}}(2005)}]{Mamon2005}%
  \BibitemOpen
  \bibfield  {author} {\bibinfo {author} {\bibfnamefont {G.~A.}\ \bibnamefont
  {{Mamon}}}\ and\ \bibinfo {author} {\bibfnamefont {E.~L.}\ \bibnamefont
  {{{\L}okas}}},\ }\href {\doibase 10.1111/j.1365-2966.2005.09400.x} {\bibfield
   {journal} {\bibinfo  {journal} {\mnras}\ }\textbf {\bibinfo {volume}
  {363}},\ \bibinfo {pages} {705} (\bibinfo {year} {2005})},\ \Eprint
  {http://arxiv.org/abs/astro-ph/0405491} {arXiv:astro-ph/0405491 [astro-ph]}
  \BibitemShut {NoStop}%
\bibitem [{\citenamefont {{Binney}}\ and\ \citenamefont
  {{Tremaine}}(2008)}]{Binney2008}%
  \BibitemOpen
  \bibfield  {author} {\bibinfo {author} {\bibfnamefont {J.}~\bibnamefont
  {{Binney}}}\ and\ \bibinfo {author} {\bibfnamefont {S.}~\bibnamefont
  {{Tremaine}}},\ }\href@noop {} {\emph {\bibinfo {title} {{Galactic Dynamics:
  Second Edition}}}}\ (\bibinfo {year} {2008})\BibitemShut {NoStop}%
\bibitem [{\citenamefont {{Mamon}}\ and\ \citenamefont
  {{Bou{\'e}}}(2010)}]{Mamon2010}%
  \BibitemOpen
  \bibfield  {author} {\bibinfo {author} {\bibfnamefont {G.~A.}\ \bibnamefont
  {{Mamon}}}\ and\ \bibinfo {author} {\bibfnamefont {G.}~\bibnamefont
  {{Bou{\'e}}}},\ }\href {\doibase 10.1111/j.1365-2966.2009.15817.x} {\bibfield
   {journal} {\bibinfo  {journal} {\mnras}\ }\textbf {\bibinfo {volume}
  {401}},\ \bibinfo {pages} {2433} (\bibinfo {year} {2010})},\ \Eprint
  {http://arxiv.org/abs/0906.4971} {arXiv:0906.4971 [astro-ph.CO]} \BibitemShut
  {NoStop}%
\bibitem [{\citenamefont {{Walker}}\ \emph
  {et~al.}(2009{\natexlab{a}})\citenamefont {{Walker}}, \citenamefont
  {{Mateo}}, \citenamefont {{Olszewski}}, \citenamefont {{Pe{\~n}arrubia}},
  \citenamefont {{Evans}},\ and\ \citenamefont {{Gilmore}}}]{Walker2009d}%
  \BibitemOpen
  \bibfield  {author} {\bibinfo {author} {\bibfnamefont {M.~G.}\ \bibnamefont
  {{Walker}}}, \bibinfo {author} {\bibfnamefont {M.}~\bibnamefont {{Mateo}}},
  \bibinfo {author} {\bibfnamefont {E.~W.}\ \bibnamefont {{Olszewski}}},
  \bibinfo {author} {\bibfnamefont {J.}~\bibnamefont {{Pe{\~n}arrubia}}},
  \bibinfo {author} {\bibfnamefont {N.~W.}\ \bibnamefont {{Evans}}}, \ and\
  \bibinfo {author} {\bibfnamefont {G.}~\bibnamefont {{Gilmore}}},\ }\href
  {\doibase 10.1088/0004-637X/704/2/1274} {\bibfield  {journal} {\bibinfo
  {journal} {\apj}\ }\textbf {\bibinfo {volume} {704}},\ \bibinfo {pages}
  {1274} (\bibinfo {year} {2009}{\natexlab{a}})},\ \Eprint
  {http://arxiv.org/abs/0906.0341} {arXiv:0906.0341 [astro-ph.CO]} \BibitemShut
  {NoStop}%
\bibitem [{\citenamefont {{Pietrzy{\'n}ski}}\ \emph {et~al.}(2009)\citenamefont
  {{Pietrzy{\'n}ski}}, \citenamefont {{G{\'o}rski}}, \citenamefont {{Gieren}},
  \citenamefont {{Ivanov}}, \citenamefont {{Bresolin}},\ and\ \citenamefont
  {{Kudritzki}}}]{Pietrzynski2009}%
  \BibitemOpen
  \bibfield  {author} {\bibinfo {author} {\bibfnamefont {G.}~\bibnamefont
  {{Pietrzy{\'n}ski}}}, \bibinfo {author} {\bibfnamefont {M.}~\bibnamefont
  {{G{\'o}rski}}}, \bibinfo {author} {\bibfnamefont {W.}~\bibnamefont
  {{Gieren}}}, \bibinfo {author} {\bibfnamefont {V.~D.}\ \bibnamefont
  {{Ivanov}}}, \bibinfo {author} {\bibfnamefont {F.}~\bibnamefont
  {{Bresolin}}}, \ and\ \bibinfo {author} {\bibfnamefont {R.-P.}\ \bibnamefont
  {{Kudritzki}}},\ }\href {\doibase 10.1088/0004-6256/138/2/459} {\bibfield
  {journal} {\bibinfo  {journal} {\aj}\ }\textbf {\bibinfo {volume} {138}},\
  \bibinfo {pages} {459} (\bibinfo {year} {2009})},\ \Eprint
  {http://arxiv.org/abs/0906.0082} {arXiv:0906.0082 [astro-ph.GA]} \BibitemShut
  {NoStop}%
\bibitem [{\citenamefont {{Irwin}}\ and\ \citenamefont
  {{Hatzidimitriou}}(1995)}]{Irwin1995}%
  \BibitemOpen
  \bibfield  {author} {\bibinfo {author} {\bibfnamefont {M.}~\bibnamefont
  {{Irwin}}}\ and\ \bibinfo {author} {\bibfnamefont {D.}~\bibnamefont
  {{Hatzidimitriou}}},\ }\href {\doibase 10.1093/mnras/277.4.1354} {\bibfield
  {journal} {\bibinfo  {journal} {\mnras}\ }\textbf {\bibinfo {volume} {277}},\
  \bibinfo {pages} {1354} (\bibinfo {year} {1995})}\BibitemShut {NoStop}%
\bibitem [{\citenamefont {{Walker}}\ \emph
  {et~al.}(2009{\natexlab{b}})\citenamefont {{Walker}}, \citenamefont
  {{Belokurov}}, \citenamefont {{Evans}}, \citenamefont {{Irwin}},
  \citenamefont {{Mateo}}, \citenamefont {{Olszewski}},\ and\ \citenamefont
  {{Gilmore}}}]{Walker2009c}%
  \BibitemOpen
  \bibfield  {author} {\bibinfo {author} {\bibfnamefont {M.~G.}\ \bibnamefont
  {{Walker}}}, \bibinfo {author} {\bibfnamefont {V.}~\bibnamefont
  {{Belokurov}}}, \bibinfo {author} {\bibfnamefont {N.~W.}\ \bibnamefont
  {{Evans}}}, \bibinfo {author} {\bibfnamefont {M.~J.}\ \bibnamefont
  {{Irwin}}}, \bibinfo {author} {\bibfnamefont {M.}~\bibnamefont {{Mateo}}},
  \bibinfo {author} {\bibfnamefont {E.~W.}\ \bibnamefont {{Olszewski}}}, \ and\
  \bibinfo {author} {\bibfnamefont {G.}~\bibnamefont {{Gilmore}}},\ }\href
  {\doibase 10.1088/0004-637X/694/2/L144} {\bibfield  {journal} {\bibinfo
  {journal} {\apjl}\ }\textbf {\bibinfo {volume} {694}},\ \bibinfo {pages}
  {L144} (\bibinfo {year} {2009}{\natexlab{b}})},\ \Eprint
  {http://arxiv.org/abs/0902.3003} {arXiv:0902.3003 [astro-ph.CO]} \BibitemShut
  {NoStop}%
\bibitem [{\citenamefont {{Fritz}}\ \emph {et~al.}(2018)\citenamefont
  {{Fritz}}, \citenamefont {{Battaglia}}, \citenamefont {{Pawlowski}},
  \citenamefont {{Kallivayalil}}, \citenamefont {{van der Marel}},
  \citenamefont {{Sohn}}, \citenamefont {{Brook}},\ and\ \citenamefont
  {{Besla}}}]{Fritz2018}%
  \BibitemOpen
  \bibfield  {author} {\bibinfo {author} {\bibfnamefont {T.~K.}\ \bibnamefont
  {{Fritz}}}, \bibinfo {author} {\bibfnamefont {G.}~\bibnamefont
  {{Battaglia}}}, \bibinfo {author} {\bibfnamefont {M.~S.}\ \bibnamefont
  {{Pawlowski}}}, \bibinfo {author} {\bibfnamefont {N.}~\bibnamefont
  {{Kallivayalil}}}, \bibinfo {author} {\bibfnamefont {R.}~\bibnamefont {{van
  der Marel}}}, \bibinfo {author} {\bibfnamefont {S.~T.}\ \bibnamefont
  {{Sohn}}}, \bibinfo {author} {\bibfnamefont {C.}~\bibnamefont {{Brook}}}, \
  and\ \bibinfo {author} {\bibfnamefont {G.}~\bibnamefont {{Besla}}},\ }\href
  {\doibase 10.1051/0004-6361/201833343} {\bibfield  {journal} {\bibinfo
  {journal} {\aap}\ }\textbf {\bibinfo {volume} {619}},\ \bibinfo {eid} {A103}
  (\bibinfo {year} {2018})},\ \Eprint {http://arxiv.org/abs/1805.00908}
  {arXiv:1805.00908 [astro-ph.GA]} \BibitemShut {NoStop}%
\bibitem [{\citenamefont {{Bonanos}}\ \emph {et~al.}(2004)\citenamefont
  {{Bonanos}}, \citenamefont {{Stanek}}, \citenamefont {{Szentgyorgyi}},
  \citenamefont {{Sasselov}},\ and\ \citenamefont {{Bakos}}}]{Bonanos2004}%
  \BibitemOpen
  \bibfield  {author} {\bibinfo {author} {\bibfnamefont {A.~Z.}\ \bibnamefont
  {{Bonanos}}}, \bibinfo {author} {\bibfnamefont {K.~Z.}\ \bibnamefont
  {{Stanek}}}, \bibinfo {author} {\bibfnamefont {A.~H.}\ \bibnamefont
  {{Szentgyorgyi}}}, \bibinfo {author} {\bibfnamefont {D.~D.}\ \bibnamefont
  {{Sasselov}}}, \ and\ \bibinfo {author} {\bibfnamefont {G.~{\'A}.}\
  \bibnamefont {{Bakos}}},\ }\href {\doibase 10.1086/381073} {\bibfield
  {journal} {\bibinfo  {journal} {\aj}\ }\textbf {\bibinfo {volume} {127}},\
  \bibinfo {pages} {861} (\bibinfo {year} {2004})},\ \Eprint
  {http://arxiv.org/abs/astro-ph/0310477} {arXiv:astro-ph/0310477 [astro-ph]}
  \BibitemShut {NoStop}%
\bibitem [{\citenamefont {{Martin}}\ \emph {et~al.}(2008)\citenamefont
  {{Martin}}, \citenamefont {{de Jong}},\ and\ \citenamefont
  {{Rix}}}]{Martin2008}%
  \BibitemOpen
  \bibfield  {author} {\bibinfo {author} {\bibfnamefont {N.~F.}\ \bibnamefont
  {{Martin}}}, \bibinfo {author} {\bibfnamefont {J.~T.~A.}\ \bibnamefont {{de
  Jong}}}, \ and\ \bibinfo {author} {\bibfnamefont {H.-W.}\ \bibnamefont
  {{Rix}}},\ }\href {\doibase 10.1086/590336} {\bibfield  {journal} {\bibinfo
  {journal} {\apj}\ }\textbf {\bibinfo {volume} {684}},\ \bibinfo {pages}
  {1075} (\bibinfo {year} {2008})},\ \Eprint {http://arxiv.org/abs/0805.2945}
  {arXiv:0805.2945 [astro-ph]} \BibitemShut {NoStop}%
\bibitem [{\citenamefont {{Walker}}\ \emph {et~al.}(2007)\citenamefont
  {{Walker}}, \citenamefont {{Mateo}}, \citenamefont {{Olszewski}},
  \citenamefont {{Gnedin}}, \citenamefont {{Wang}}, \citenamefont {{Sen}},\
  and\ \citenamefont {{Woodroofe}}}]{Walker2007}%
  \BibitemOpen
  \bibfield  {author} {\bibinfo {author} {\bibfnamefont {M.~G.}\ \bibnamefont
  {{Walker}}}, \bibinfo {author} {\bibfnamefont {M.}~\bibnamefont {{Mateo}}},
  \bibinfo {author} {\bibfnamefont {E.~W.}\ \bibnamefont {{Olszewski}}},
  \bibinfo {author} {\bibfnamefont {O.~Y.}\ \bibnamefont {{Gnedin}}}, \bibinfo
  {author} {\bibfnamefont {X.}~\bibnamefont {{Wang}}}, \bibinfo {author}
  {\bibfnamefont {B.}~\bibnamefont {{Sen}}}, \ and\ \bibinfo {author}
  {\bibfnamefont {M.}~\bibnamefont {{Woodroofe}}},\ }\href {\doibase
  10.1086/521998} {\bibfield  {journal} {\bibinfo  {journal} {\apjl}\ }\textbf
  {\bibinfo {volume} {667}},\ \bibinfo {pages} {L53} (\bibinfo {year}
  {2007})},\ \Eprint {http://arxiv.org/abs/0708.0010} {arXiv:0708.0010
  [astro-ph]} \BibitemShut {NoStop}%
\bibitem [{\citenamefont {{Bellazzini}}\ \emph {et~al.}(2004)\citenamefont
  {{Bellazzini}}, \citenamefont {{Gennari}}, \citenamefont {{Ferraro}},\ and\
  \citenamefont {{Sollima}}}]{Bellazzini2004}%
  \BibitemOpen
  \bibfield  {author} {\bibinfo {author} {\bibfnamefont {M.}~\bibnamefont
  {{Bellazzini}}}, \bibinfo {author} {\bibfnamefont {N.}~\bibnamefont
  {{Gennari}}}, \bibinfo {author} {\bibfnamefont {F.~R.}\ \bibnamefont
  {{Ferraro}}}, \ and\ \bibinfo {author} {\bibfnamefont {A.}~\bibnamefont
  {{Sollima}}},\ }\href {\doibase 10.1111/j.1365-2966.2004.08226.x} {\bibfield
  {journal} {\bibinfo  {journal} {\mnras}\ }\textbf {\bibinfo {volume} {354}},\
  \bibinfo {pages} {708} (\bibinfo {year} {2004})},\ \Eprint
  {http://arxiv.org/abs/astro-ph/0407444} {arXiv:astro-ph/0407444 [astro-ph]}
  \BibitemShut {NoStop}%
\bibitem [{\citenamefont {{Mateo}}\ \emph {et~al.}(2008)\citenamefont
  {{Mateo}}, \citenamefont {{Olszewski}},\ and\ \citenamefont
  {{Walker}}}]{Mateo2008}%
  \BibitemOpen
  \bibfield  {author} {\bibinfo {author} {\bibfnamefont {M.}~\bibnamefont
  {{Mateo}}}, \bibinfo {author} {\bibfnamefont {E.~W.}\ \bibnamefont
  {{Olszewski}}}, \ and\ \bibinfo {author} {\bibfnamefont {M.~G.}\ \bibnamefont
  {{Walker}}},\ }\href {\doibase 10.1086/522326} {\bibfield  {journal}
  {\bibinfo  {journal} {\apj}\ }\textbf {\bibinfo {volume} {675}},\ \bibinfo
  {pages} {201} (\bibinfo {year} {2008})},\ \Eprint
  {http://arxiv.org/abs/0708.1327} {arXiv:0708.1327 [astro-ph]} \BibitemShut
  {NoStop}%
\bibitem [{\citenamefont {{Bellazzini}}\ \emph {et~al.}(2005)\citenamefont
  {{Bellazzini}}, \citenamefont {{Gennari}},\ and\ \citenamefont
  {{Ferraro}}}]{Bellazzini2005}%
  \BibitemOpen
  \bibfield  {author} {\bibinfo {author} {\bibfnamefont {M.}~\bibnamefont
  {{Bellazzini}}}, \bibinfo {author} {\bibfnamefont {N.}~\bibnamefont
  {{Gennari}}}, \ and\ \bibinfo {author} {\bibfnamefont {F.~R.}\ \bibnamefont
  {{Ferraro}}},\ }\href {\doibase 10.1111/j.1365-2966.2005.09027.x} {\bibfield
  {journal} {\bibinfo  {journal} {\mnras}\ }\textbf {\bibinfo {volume} {360}},\
  \bibinfo {pages} {185} (\bibinfo {year} {2005})},\ \Eprint
  {http://arxiv.org/abs/astro-ph/0503418} {arXiv:astro-ph/0503418 [astro-ph]}
  \BibitemShut {NoStop}%
\bibitem [{\citenamefont {{Koch}}\ \emph {et~al.}(2007)\citenamefont {{Koch}},
  \citenamefont {{Kleyna}}, \citenamefont {{Wilkinson}}, \citenamefont
  {{Grebel}}, \citenamefont {{Gilmore}}, \citenamefont {{Evans}}, \citenamefont
  {{Wyse}},\ and\ \citenamefont {{Harbeck}}}]{Koch2007}%
  \BibitemOpen
  \bibfield  {author} {\bibinfo {author} {\bibfnamefont {A.}~\bibnamefont
  {{Koch}}}, \bibinfo {author} {\bibfnamefont {J.~T.}\ \bibnamefont
  {{Kleyna}}}, \bibinfo {author} {\bibfnamefont {M.~I.}\ \bibnamefont
  {{Wilkinson}}}, \bibinfo {author} {\bibfnamefont {E.~K.}\ \bibnamefont
  {{Grebel}}}, \bibinfo {author} {\bibfnamefont {G.~F.}\ \bibnamefont
  {{Gilmore}}}, \bibinfo {author} {\bibfnamefont {N.~W.}\ \bibnamefont
  {{Evans}}}, \bibinfo {author} {\bibfnamefont {R.~F.~G.}\ \bibnamefont
  {{Wyse}}}, \ and\ \bibinfo {author} {\bibfnamefont {D.~R.}\ \bibnamefont
  {{Harbeck}}},\ }\href {\doibase 10.1086/519380} {\bibfield  {journal}
  {\bibinfo  {journal} {\aj}\ }\textbf {\bibinfo {volume} {134}},\ \bibinfo
  {pages} {566} (\bibinfo {year} {2007})},\ \Eprint
  {http://arxiv.org/abs/0704.3437} {arXiv:0704.3437 [astro-ph]} \BibitemShut
  {NoStop}%
\bibitem [{\citenamefont {{Pietrzy{\'n}ski}}\ \emph {et~al.}(2008)\citenamefont
  {{Pietrzy{\'n}ski}}, \citenamefont {{Gieren}}, \citenamefont {{Szewczyk}},
  \citenamefont {{Walker}}, \citenamefont {{Rizzi}}, \citenamefont
  {{Bresolin}}, \citenamefont {{Kudritzki}}, \citenamefont {{Nalewajko}},
  \citenamefont {{Storm}}, \citenamefont {{Dall'Ora}},\ and\ \citenamefont
  {{Ivanov}}}]{Pietrzynski2008}%
  \BibitemOpen
  \bibfield  {author} {\bibinfo {author} {\bibfnamefont {G.}~\bibnamefont
  {{Pietrzy{\'n}ski}}}, \bibinfo {author} {\bibfnamefont {W.}~\bibnamefont
  {{Gieren}}}, \bibinfo {author} {\bibfnamefont {O.}~\bibnamefont
  {{Szewczyk}}}, \bibinfo {author} {\bibfnamefont {A.}~\bibnamefont
  {{Walker}}}, \bibinfo {author} {\bibfnamefont {L.}~\bibnamefont {{Rizzi}}},
  \bibinfo {author} {\bibfnamefont {F.}~\bibnamefont {{Bresolin}}}, \bibinfo
  {author} {\bibfnamefont {R.-P.}\ \bibnamefont {{Kudritzki}}}, \bibinfo
  {author} {\bibfnamefont {K.}~\bibnamefont {{Nalewajko}}}, \bibinfo {author}
  {\bibfnamefont {J.}~\bibnamefont {{Storm}}}, \bibinfo {author} {\bibfnamefont
  {M.}~\bibnamefont {{Dall'Ora}}}, \ and\ \bibinfo {author} {\bibfnamefont
  {V.}~\bibnamefont {{Ivanov}}},\ }\href {\doibase
  10.1088/0004-6256/135/6/1993} {\bibfield  {journal} {\bibinfo  {journal}
  {\aj}\ }\textbf {\bibinfo {volume} {135}},\ \bibinfo {pages} {1993} (\bibinfo
  {year} {2008})},\ \Eprint {http://arxiv.org/abs/0804.0347} {arXiv:0804.0347
  [astro-ph]} \BibitemShut {NoStop}%
\bibitem [{\citenamefont {{Lee}}\ \emph {et~al.}(2009)\citenamefont {{Lee}},
  \citenamefont {{Yuk}}, \citenamefont {{Park}}, \citenamefont {{Harris}},\
  and\ \citenamefont {{Zaritsky}}}]{Lee2009}%
  \BibitemOpen
  \bibfield  {author} {\bibinfo {author} {\bibfnamefont {M.~G.}\ \bibnamefont
  {{Lee}}}, \bibinfo {author} {\bibfnamefont {I.-S.}\ \bibnamefont {{Yuk}}},
  \bibinfo {author} {\bibfnamefont {H.~S.}\ \bibnamefont {{Park}}}, \bibinfo
  {author} {\bibfnamefont {J.}~\bibnamefont {{Harris}}}, \ and\ \bibinfo
  {author} {\bibfnamefont {D.}~\bibnamefont {{Zaritsky}}},\ }\href {\doibase
  10.1088/0004-637X/703/1/692} {\bibfield  {journal} {\bibinfo  {journal}
  {\apj}\ }\textbf {\bibinfo {volume} {703}},\ \bibinfo {pages} {692} (\bibinfo
  {year} {2009})},\ \Eprint {http://arxiv.org/abs/0907.5102} {arXiv:0907.5102
  [astro-ph.CO]} \BibitemShut {NoStop}%
\bibitem [{\citenamefont {{Carrera}}\ \emph {et~al.}(2002)\citenamefont
  {{Carrera}}, \citenamefont {{Aparicio}}, \citenamefont
  {{Mart{\'\i}nez-Delgado}},\ and\ \citenamefont
  {{Alonso-Garc{\'\i}a}}}]{Carrera2002}%
  \BibitemOpen
  \bibfield  {author} {\bibinfo {author} {\bibfnamefont {R.}~\bibnamefont
  {{Carrera}}}, \bibinfo {author} {\bibfnamefont {A.}~\bibnamefont
  {{Aparicio}}}, \bibinfo {author} {\bibfnamefont {D.}~\bibnamefont
  {{Mart{\'\i}nez-Delgado}}}, \ and\ \bibinfo {author} {\bibfnamefont
  {J.}~\bibnamefont {{Alonso-Garc{\'\i}a}}},\ }\href {\doibase 10.1086/340702}
  {\bibfield  {journal} {\bibinfo  {journal} {\aj}\ }\textbf {\bibinfo {volume}
  {123}},\ \bibinfo {pages} {3199} (\bibinfo {year} {2002})},\ \Eprint
  {http://arxiv.org/abs/astro-ph/0203300} {arXiv:astro-ph/0203300 [astro-ph]}
  \BibitemShut {NoStop}%
\bibitem [{\citenamefont {{Walker}}\ \emph
  {et~al.}(2009{\natexlab{c}})\citenamefont {{Walker}}, \citenamefont
  {{Mateo}}, \citenamefont {{Olszewski}}, \citenamefont {{Sen}},\ and\
  \citenamefont {{Woodroofe}}}]{Walker2009b}%
  \BibitemOpen
  \bibfield  {author} {\bibinfo {author} {\bibfnamefont {M.~G.}\ \bibnamefont
  {{Walker}}}, \bibinfo {author} {\bibfnamefont {M.}~\bibnamefont {{Mateo}}},
  \bibinfo {author} {\bibfnamefont {E.~W.}\ \bibnamefont {{Olszewski}}},
  \bibinfo {author} {\bibfnamefont {B.}~\bibnamefont {{Sen}}}, \ and\ \bibinfo
  {author} {\bibfnamefont {M.}~\bibnamefont {{Woodroofe}}},\ }\href {\doibase
  10.1088/0004-6256/137/2/3109} {\bibfield  {journal} {\bibinfo  {journal}
  {\aj}\ }\textbf {\bibinfo {volume} {137}},\ \bibinfo {pages} {3109} (\bibinfo
  {year} {2009}{\natexlab{c}})},\ \Eprint {http://arxiv.org/abs/0811.1990}
  {arXiv:0811.1990 [astro-ph]} \BibitemShut {NoStop}%
\bibitem [{\citenamefont {{Walker}}\ \emph
  {et~al.}(2009{\natexlab{d}})\citenamefont {{Walker}}, \citenamefont
  {{Mateo}},\ and\ \citenamefont {{Olszewski}}}]{Walker2009a}%
  \BibitemOpen
  \bibfield  {author} {\bibinfo {author} {\bibfnamefont {M.~G.}\ \bibnamefont
  {{Walker}}}, \bibinfo {author} {\bibfnamefont {M.}~\bibnamefont {{Mateo}}}, \
  and\ \bibinfo {author} {\bibfnamefont {E.~W.}\ \bibnamefont {{Olszewski}}},\
  }\href {\doibase 10.1088/0004-6256/137/2/3100} {\bibfield  {journal}
  {\bibinfo  {journal} {\aj}\ }\textbf {\bibinfo {volume} {137}},\ \bibinfo
  {pages} {3100} (\bibinfo {year} {2009}{\natexlab{d}})},\ \Eprint
  {http://arxiv.org/abs/0811.0118} {arXiv:0811.0118 [astro-ph]} \BibitemShut
  {NoStop}%
\bibitem [{\citenamefont {{Foreman-Mackey}}\ \emph {et~al.}(2013)\citenamefont
  {{Foreman-Mackey}}, \citenamefont {{Hogg}}, \citenamefont {{Lang}},\ and\
  \citenamefont {{Goodman}}}]{emcee}%
  \BibitemOpen
  \bibfield  {author} {\bibinfo {author} {\bibfnamefont {D.}~\bibnamefont
  {{Foreman-Mackey}}}, \bibinfo {author} {\bibfnamefont {D.~W.}\ \bibnamefont
  {{Hogg}}}, \bibinfo {author} {\bibfnamefont {D.}~\bibnamefont {{Lang}}}, \
  and\ \bibinfo {author} {\bibfnamefont {J.}~\bibnamefont {{Goodman}}},\ }\href
  {\doibase 10.1086/670067} {\bibfield  {journal} {\bibinfo  {journal}
  {Publications of the Astronomical Society of the Pacific}\ }\textbf {\bibinfo
  {volume} {125}},\ \bibinfo {pages} {306} (\bibinfo {year} {2013})},\ \Eprint
  {http://arxiv.org/abs/1202.3665} {arXiv:1202.3665 [astro-ph.IM]} \BibitemShut
  {NoStop}%
\bibitem [{\citenamefont {{de Martino}}\ \emph {et~al.}(2022)\citenamefont {{de
  Martino}}, \citenamefont {{Diaferio}},\ and\ \citenamefont
  {{Ostorero}}}]{deMartino2022}%
  \BibitemOpen
  \bibfield  {author} {\bibinfo {author} {\bibfnamefont {I.}~\bibnamefont {{de
  Martino}}}, \bibinfo {author} {\bibfnamefont {A.}~\bibnamefont {{Diaferio}}},
  \ and\ \bibinfo {author} {\bibfnamefont {L.}~\bibnamefont {{Ostorero}}},\
  }\href {\doibase 10.1093/mnras/stac2336} {\bibfield  {journal} {\bibinfo
  {journal} {\mnras}\ }\textbf {\bibinfo {volume} {516}},\ \bibinfo {pages}
  {3556} (\bibinfo {year} {2022})},\ \Eprint {http://arxiv.org/abs/2208.14110}
  {arXiv:2208.14110 [astro-ph.GA]} \BibitemShut {NoStop}%
\bibitem [{\citenamefont {{The Theia Collaboration}}(2017)}]{Theia2017}%
  \BibitemOpen
  \bibfield  {author} {\bibinfo {author} {\bibnamefont {{The Theia
  Collaboration}}},\ }\href {\doibase 10.48550/arXiv.1707.01348} {\bibfield
  {journal} {\bibinfo  {journal} {arXiv e-prints}\ ,\ \bibinfo {eid}
  {arXiv:1707.01348}} (\bibinfo {year} {2017})},\ \Eprint
  {http://arxiv.org/abs/1707.01348} {arXiv:1707.01348 [astro-ph.IM]}
  \BibitemShut {NoStop}%
\bibitem [{\citenamefont {{Malbet}}\ and\ \citenamefont {{et
  al.}}(2019)}]{Malbet2019}%
  \BibitemOpen
  \bibfield  {author} {\bibinfo {author} {\bibfnamefont {F.}~\bibnamefont
  {{Malbet}}}\ and\ \bibinfo {author} {\bibnamefont {{et al.}}},\ }\href
  {\doibase 10.48550/arXiv.1910.08028} {\bibfield  {journal} {\bibinfo
  {journal} {arXiv e-prints}\ ,\ \bibinfo {eid} {arXiv:1910.08028}} (\bibinfo
  {year} {2019})},\ \Eprint {http://arxiv.org/abs/1910.08028} {arXiv:1910.08028
  [astro-ph.IM]} \BibitemShut {NoStop}%
\bibitem [{\citenamefont {{Malbet}}\ and\ \citenamefont {{et
  al.}}(2021)}]{Malbet2021}%
  \BibitemOpen
  \bibfield  {author} {\bibinfo {author} {\bibfnamefont {F.}~\bibnamefont
  {{Malbet}}}\ and\ \bibinfo {author} {\bibnamefont {{et al.}}},\ }\href
  {\doibase 10.1007/s10686-021-09781-1} {\bibfield  {journal} {\bibinfo
  {journal} {Experimental Astronomy}\ }\textbf {\bibinfo {volume} {51}},\
  \bibinfo {pages} {845} (\bibinfo {year} {2021})},\ \Eprint
  {http://arxiv.org/abs/2111.08709} {arXiv:2111.08709 [astro-ph.IM]}
  \BibitemShut {NoStop}%
\end{thebibliography}%

\end{document}